\documentclass[11pt]{article}
\usepackage{CJK}
\usepackage{graphicx}
\usepackage{fancyhdr}
\usepackage{amsmath,amsthm,amssymb}
\usepackage{graphicx,color,epstopdf}
\usepackage{mathrsfs}
\usepackage{multirow}
\usepackage{indentfirst}
\usepackage{color}
\renewcommand{\baselinestretch}{1.15}
\newtheorem{theorem}{Theorem}
\newtheorem{remark}{Remark}
\newtheorem{lemma}{Lemma}

\makeatletter
\newcommand{\Rmnum}[1]{\expandafter\@slowromancap\romannumeral #1@}
\makeatother
\newtheorem{corollary}{Corollary}
\allowdisplaybreaks
\def\hDash{\bot\!\!\!\bot}
\begin{document}
\date{}

\title
{\Large \bf A robust  adaptive-to-model enhancement test for parametric single-index models} {\small
\author{Cuizhen Niu$^1$ and Lixing Zhu$^2$\footnote{Corresponding author: Lixing Zhu, email: lzhu@hkbu.edu.hk. The research described herewith was supported by a grant from the University Grants Council of Hong Kong and the Outstanding Innovative Talents Cultivation Funded Programs 2014 of Renmin University of China.}\\
{\small {\small {\it $^1$School of Statistics, Renmin
University of China, Beijing } }}\\
{\small {\small {\it $^2$Department of Mathematics, Hong Kong Baptist
University, Hong Kong } }}
 }}
\date{}
\maketitle

\renewcommand\baselinestretch{1.4}
{\small

\noindent {\bf Abstract:}   In the research on checking whether the underlying model is  of parametric single-index structure with outliers in observations, the purpose of this paper is two-fold. First, a test that is robust against outliers is suggested. The Hampel's second-order influence function of the test statistic is proved to be bounded. Second, the test fully uses the dimension reduction structure of the hypothetical model and automatically adapts to alternative models when the null hypothesis is false. Thus, the test can greatly overcome the dimensionality problem and is still  omnibus against general alternative models.   The performance of the test is demonstrated by both Monte Carlo simulation studies and an application to a real dataset.

\bigskip

}
\newpage

\section{Introduction}
\renewcommand{\theequation}{1.\arabic{equation}}
\setcounter{equation}{0}
Testing the validity of a specified model is an important issue in statistical inference. A long-standing focus for this model checking problem is their sensitivity to outlying observations or heavy-tailed distributions, which may have destructive effects even though with small violation in usual observations. However, in the past decades, researchers have made more effort on robust estimation whereas paid less attention to robust hypothesis testing. Therefore, it is critical to develop a robust  test that can be against outlier contamination.

Outliers or contamination data are a ubiquitous problem in many disciplines, for example, clinical trials, medical research, longitudinal studies and so forth. When there exist outliers in the data, robust statistical inference procedures are necessary to improve the accuracy and reliability of results. The purpose of robust testing is two-fold, just as stated in Heritier and Ronchetti (1994): One is that under small and arbitrary departures from the null hypothesis, the level of a test should be stable, which is called the robustness of validity; The other is that the test can still make a good power performance under small and arbitrary departures from specified alternatives, that is called the robustness of efficiency. Wang and Qu (2007) suggested a robust version of Zheng's  (1996) test. Their numerical studies also showed the necessity of using a robust testing procedure: the effect of outliers on Zheng's original test is dramatic and destructive so that it can not maintain the significance level.

Many efforts have been  devoted to the development of robust testing procedures. For linear regression models, Schrader and Hettmansperger (1980) proposed the $\rho_c$ test based on Huber's M estimatiors; Markatou and Hettmansperger (1990) introduced an aligned generalized M test for testing subhypotheses in general linear models, which is a robustification of the well known $F$ test and can be viewed as a generalization of Sen's (1982) $M$ test for linear models. Afterwards, Heritier and Ronchetti (1994) and Markatou and Manos (1996) presented robust versions of the Wald, score and drop-in-dispersion tests for general parametric models and nonlinear regression models, respectively. Wang and Qu (2007) developed a robust approach for testing the parametric form of a regression function versus an omnibus alternative, which can be viewed as a robustification of a smoothing-based conditional moment test. Feng et al (2015) recommended a robust testing procedure to make comparison of two regression curves through combining a Wilcoxon-type artificial likelihood function with generalized likelihood ratio test.

There are a number of proposals available in the literature on testing consistently the correct specification of a parametric regression model. Most of  existing test procedures can be classified into two categories: global smoothing tests and local smoothing tests. As mentioned in a comprehensive review paper of Gonz\'alez-Manteiga and Crujeiras (2013), the global smoothing tests mainly involve empirical process, which can avoid subjective selection of the smoothing parameter, such as bandwidth; The local smoothing tests are based on nonparametric smoothing techniques such as Nadaraya-Waston kernel estimation (Nadaraya 1964; Waston 1964), smoothing spline estimation or other local smoothing techniques. Examples of global smoothing tests include Bierens (1990), Bierens and Ploberger (1997), Stute (1997), Stute et al (1998b), Whang (2000), Escanciano (2006a), among many others. This class of methods enjoy fast convergence rate of order $O(n^{-1/2})$ (see Stute et al (1998a)). However, in high-dimensional scenarios, the limiting null distribution is usually intractable which requires the assistance from resampling approximation to determine critical values and is not sensitive to high-frequency models. Further, the power performance in high-dimensional cases is not very encouraging.  This problem makes greatly practical limitation since it is not uncommon to have high-order or high-frequency models. 

As for local smoothing tests, examples include the tests suggested by H\"ardle and Mammen (1993), Zheng (1996), Fan et al (2001), Horowitz and Spokoiny (2001), Koul and Ni (2004), Van Keilegom et al (2008). Since they must involve multivariate nonparametric function estimation procedures, and thus inevitably suffer from the curse of dimensionality  when the number of covariates is large, even moderate. This typical  problem is a big obstacle for the tests to well maintain the significance level and sense the alternative models. Because of the data sparseness in multidimensional spaces, the behavior of nonparametric smooth estimators quickly deteriorates as the dimension increases, see Stone (1980). Further,  even when there are no outliers, local smoothing tests  have the typical  convergence rate of order $O(n^{-1/2}h^{-p/4})$ to their limits, which is very slow when $p$ is large where $p$ is the number of covariates. Besides, a suitable choice of smooth parameter is difficult but necessary for these tests. Although previous simulation and empirical studies show that the effect of bandwidth selection is not too profound for small $p$ situation, how to make an optimal bandwidth choice is not solved thoroughly when the dimension $p$ is relatively large.

The above analysis shows that there exists a common problem in both procedures, the global and the local, that is, the sparseness of data in high-dimensional spaces makes most of test statistics suffer curse of dimensionality, even for large sample sizes (see Escanciano (2007)). To attack this challenge, a representative method documented as projection-pursuit technique was proposed and experimented. The significant feature of this method is to employ the projection of original data onto one-dimensional subspaces: first projecting the original high-dimensional covariates to one-dimensional space to form a linear combination and a test can be obtained as an average of tests based on these selected combinations, see Huber (1985) for detail.
Escanciano (2006b) proposed a consistent test for the goodness-of-fit of parametric regression models, which applied a residual marked empirical process based on projections to bypass the curse of dimensionality caused by the fact that high-dimensional space is mostly empty. Zhu and Li (1998) suggested to use projection pursuit technique to define a test that is based on   an unweighed integral of expectations with respect to all one-dimensional directions. Zhu and An (1992) had already used this idea to deal with a relevant testing problem. Zhu (2003) constructed a lack-of-fit test via seeking for a good projection direction for plotting to achieve the dimension-reduction aim.  Lavergne and Patilea (2008) introduced the projection-pursuit technique to local smoothing-based tests to avoid the effect of dimension.  Afterwards, Lavergne and Patilea (2012) suggested a smooth integrated conditional moment (ICM) test, which is an omnibus test based on the kernel estimation that performs against a sequence of directional nonparametric alternatives as if there were only one regressor whatever the number of regressors. However, all of tests require resampling/bootstrap to determine the critical values, which is compute-intensive and time-consuming. Stute and Zhu (2002) simply used the one-dimensional projected covariate that is based on the null model and thus the problem of dimensionality is greatly alleviated. But the disadvantage is that it is a directional rather than an omnibus test which cannot detect general alternatives. Another relevant reference is Stute, Xu and Zhu (2008) who also suggested a dimension reduction test that is based on the residual empirical process marked by a set of functions of the covariates. This test relies solely on selecting proper functions for the significance level maintainance and power enhancement.

Recently, a dimension-reduction model-adaptive test is proposed by Guo et al (2015), which is an omnibus test against global alternative models. This test introduces a model-adaptation concept in model checking for parametric regression models. The test statistic under the hypothetical model can converge to its limit at  the rate of order $O(n^{-1/2}h^{-1/4})$ and detect local alternatives distinct from the null model at this rate, which is not affected by the dimension of covariates. Their test behaves like a local smoothing test, as if the covariates were one-dimensional. Another superiority is that it owns tractable limiting null distribution and can work very well even with moderate sample sizes without the assistance of resampling approximation to determine critical values.

All of the above tests can avoid the curse of dimensionality to some extent, however, they are not robust against outliers and their efficiency is adversely affected by outlying observations.  Our subsequent numerical analysis suggests that the test proposed by Guo et al (2015) is failure when there exist some outliers because a linear local average of the response variable is not robust, as elaborated in H\"ardle (1992). To address this problem, we intend to incorporate the idea in Guo et al (2015) into our robust model-adaptive smoothing-based conditional moment test so that it can possess the robustness property and simultaneously solve the problem of dimensionality.

The hypothetical model is  the following with a dimension-reduction structure:
\begin{equation}\label{null_model}
    Y=g(\beta^\top X,\theta)+e,
\end{equation}
where $Y$ is the response with the covariate $X\in \mathbb{R}^p$, the error $e$ is with zero mean and  is independent of $X$. $\beta$ and $\theta$ are unknown parameter vectors of dimensions $p$ and $d$, respectively. In addition, $g(\cdot)$ is a known function and the superscript $\top$ denotes transposition. As we often have no much information in advance on model structure under alternative hypothesis, a  general alternative model is considered as follows:
\begin{equation}\label{alter_model}
    Y=m(B^\top X)+\varepsilon,
\end{equation}
where $m(\cdot)$ is an unknown smooth function and $E(\varepsilon|X)=0$. $B$ is a $p\times q$ matrix with $q$ orthogonal columns for an unknown number $q$ with $1\leq q\leq p$. This model treats the nonparametric regression $E(Y|X)=m(X)$ as a special case for which the matrix $B=I_p$ with $q=p$: 
$$Y=m(X)+\varepsilon,$$
where $m(\cdot)$ is an unknown smooth link function and $E(\varepsilon|X)=0$.

In this paper, we construct a robust dimension reduction adaptive-to-model test (RDREAM). It sufficiently invokes the information in both the null and alternative models to get rid of curse of dimensionality and employs centered asymptotic rank transformation technique to achieve the goal of robustness.
 We further study the local robustness via influence function analysis, which indicates that our RDREAM has first-order influence function of zero and second-order influence function bounded in the response direction. Therefore, it is verified that our test can make more stable and robust performance when there are outliers in responses than existing local smoothing tests.

The rest of this article is organized as follows. In Section~2, the test is constructed. The approaches to estimate the matrix $B$ and to identify its structure dimension $q$ are also stated in this section. Section~3 presents the  large sample properties under the null, global and local alternative hypothesis. Section~4 discusses the local robustness property through the Hampel influence function analysis. Numerical studies including simulation studies and a HIV real data analysis are respectively reported in Section~5 and Section~6. All of the proofs are relegated to the Appendix.

\section{A robust dimension reduction adaptive-to-model test}\label{sec2}
\renewcommand{\theequation}{2.\arabic{equation}}
\setcounter{equation}{0}

As discussed before, the hypotheses of interest are:
\begin{eqnarray}
  &&H_0: E(Y|X)=g(\beta^\top X,\theta)~\mbox{for some}~\beta\in R^p, \theta\in R^d;\nonumber\\
  &&H_1: E(Y|X)=m(B^\top X)\neq g(\beta^\top X,\theta)~\mbox{for any}~\beta\in R^p, \theta\in R^d,~~\label{hypothesis}
\end{eqnarray}
where $g(\cdot)$ is a known link function. $\beta$ and $\theta$ are respectively the parameter vectors of $p$ and $d$ dimensions. $B$ is a $p\times q$ orthonormal matrix where $B^\top B=I_q$ and $1\leq q\leq p$.

\subsection{Test statistic construction}
The key idea of the local smoothing-based conditional moment test is to apply the centered rank-transformed residuals. Denote $e=Y-g(\beta^\top X,\theta)$
and let $e^\star_i=H(e_i)-\frac{1}{2}$, where $H(\cdot)$ is the distribution function of $e$ and $H(e_i),\,i=1,\ldots,n$ follow
a uniform distribution on $(0,1)$.
Under the null hypothesis $H_0$,
\begin{eqnarray}\label{expect_condition}
  E(e^\star_i|X_i)&=&E\Big\{H(e_i)-\frac{1}{2}|X_i\Big\}=0,
\end{eqnarray}
 In this case,  the model (\ref{null_model}) is with $q=1$, and
\begin{equation*}
    E(e^\star_i|X_i)=0= E(e^\star_i|\beta^\top X_i)=E(e^\star_i|B^\top X_i)=0.
\end{equation*}
Further, the following formula
\begin{equation}\label{mean_H0}
    E\{e^\star_iE(e^\star_i|B^\top X_i)f(B^\top X_i)\}=E\{E^2(e^\star_i|B^\top X_i)f(B^\top X_i)\}=0
\end{equation}
holds under $H_0$, where $f(\cdot)$ is the probability density function of $B^\top X_i$.

Under  $H_1$, we have $e=Y-g(\beta^\top X,\theta)=m(B^\top X)-g(\beta^\top X,\theta)+\varepsilon$. It is easy to see that \begin{eqnarray}
 && E(e^\star_i|X_i)\nonumber\\
 &=&E\Big\{Q\Big(\varepsilon_i +(m(B^\top X_i)-g(\beta^\top X_i,\theta))-(m(B^\top X)-g(\beta^\top X,\theta))\Big)-\frac{1}{2}|X_i\Big\},~~~~~~~\label{expect_H1}
\end{eqnarray}
where $Q(\cdot)$ is the distribution function of $\varepsilon$. The conditional expectation $E(e_i^\star|X_i)$ equals to zero only if for any $x$ in a set with probability $1$, there is
\begin{equation}\label{exeqzero}
    E\Big\{Q\Big(\varepsilon_i +(m(B^\top x)-g(\beta^\top x,\theta))-(m(B^\top X)-g(\beta^\top X,\theta))\Big)-\frac{1}{2}\Big\}=0.
\end{equation}
It occurs only when $P(m(B^\top X_j)-g(\beta^\top X_j,\theta)=C)=1$ for some constant $C$, which only holds under the null hypothesis $H_0$. Since we can enlarge the null class of models by including some location shifts, if $g(\beta^\top x,\theta)$ belongs to the null class of models, then so does $g(\beta^\top x,\theta)+C$; in other words, it is reasonable to assume that the null class of models is sufficiently general to contain all location shifts in the $y$ direction. Therefore, under $H_1$, the formula (\ref{exeqzero}) does not hold. Further, based on
(\ref{expect_H1}), the conditional expectation $E(e_i^\star|X_i)$ is not zero and
\begin{equation*}
    E(e_i^\star|X_i)\neq 0\Leftrightarrow E(e_i^\star|B^\top X_i)\neq 0.
\end{equation*}
Thus, we have
\begin{equation}\label{mean_H1}
    E\{e^\star_iE(e^\star_i|B^\top X_i)f(B^\top X_i)\}=E\{E^2(e^\star_i|B^\top X_i)f(B^\top X_i)\}>0.
\end{equation}

Based on the different performance of $E\{e^\star_iE(e^\star_i|B^\top X_i)f(B^\top X_i)\}$ under $H_0$ and $H_1$ in (\ref{mean_H0}) and (\ref{mean_H1}), respectively, the empirical version of it can be applied to construct a test statistic.
The null hypothesis $H_0$ is rejected for large values of the test statistic.

Given a random sample $\{(y_1,x_1), (y_2,x_2),\cdots,(y_n,x_n)\}$, define the asymptotic rank transform of $\hat e_i=y_i-g(\hat\beta^\top x_i,\hat\theta)$
as $n^{-1}\sum_{l=1}^n I(\hat e_l\leq\hat e_i)$ where $\sum_{l=1}^n I(\hat e_l\leq\hat e_i)$ is the rank of $\hat e_i$ among all of the $n$ residuals. Here, $\hat\beta$ and $\hat\theta$ come from robust etimates of $\beta$ and $\theta$. Further, the corresponding centered asymptotic rank transform of residuals is as follows
\begin{equation}\label{e_star}
    \hat e_i^\star=\frac{1}{n}\sum_{l=1}^n I(\hat e_l\leq\hat e_i)-\frac{n+1}{2n},~~l=1,\ldots,n,
\end{equation}
where $I(\cdot)$ is the indicator function.

Once an estimator of $\hat B(\hat q)$ is available, a kernel estimator of the regression function $E(e^\star_i|B^\top X_i)$ can be estimated as follows:
\begin{equation*}
    \hat E(e^\star_i|\hat B(\hat q)^\top x_i)=\frac{\frac{1}{n-1}\sum_{j\neq i}^n \hat e_j^\star \mathcal{K}_h\{\hat B(\hat q)^\top (x_i-x_j)\}}{\hat f(\hat B(\hat q)^\top x_i)},
\end{equation*}
where $\hat e_j^\star$ has been defined in (\ref{e_star}). $\hat B(\hat q)$ is an sufficient dimension reduction (SDR) estimate of the matrix $B$ with an estimated structural dimension $\hat q$ of $q$ and the estimates will be specified later. Besides, $\mathcal{K}_h(\cdot)=\mathcal{K}(\cdot/h)/h^{\hat q}$, where
$\mathcal{K}(\cdot)$ is a $\hat q$-dimensional kernel function and $h$ is the bandwidth,
and $\hat f(\hat B(\hat q)^\top x_i)$ is a kernel estimator of the density function of $f(B^\top x_i)$,
\begin{equation*}
    \hat f(\hat B(\hat q)^\top x_i)=\frac{1}{n-1}\sum_{j\neq i}^n  \mathcal{K}_h\{\hat B(\hat q)^\top (x_i-x_j)\}.
\end{equation*}
Further, a robust dimension reduction adaptive-to-model test (RDREAM) can be constructed as follows:
\begin{equation}\label{statistic_Vn}
    V_n=\frac{1}{n(n-1)} \sum_{i=1}^n\sum_{j\neq i}^n \mathcal{K}_h\{\hat B(\hat q)^\top (x_i-x_j)\}\hat e_i^\star\hat e_j^\star.
\end{equation}

\begin{remark}
From the above construction of $V_n$ in (\ref{statistic_Vn}), it seems that except for the estimates of the matrix $B$ and structural dimension $q$, the test statistic makes no difference with the test proposed by Wang and Qu (2007) as follows:
\begin{equation}\label{statistic_Vnwan}
    \tilde{V}_n=\frac{1}{n(n-1)} \sum_{i=1}^n\sum_{j\neq i}^n \tilde{\mathcal{K}}_h (x_i-x_j)\hat e_i^\star\hat e_j^\star,
\end{equation}
where $\tilde{\mathcal{K}}_h(\cdot)=\tilde{\mathcal{K}}(\cdot/h)/h^p$ with $\tilde{\mathcal{K}}(\cdot)$ being a $p$-dimensional kernel function. Comparing the test statistic in (\ref{statistic_Vn}) with that in (\ref{statistic_Vnwan}), we note that  $\hat q$-dimensional kernel function ($\hat q\leq p$) is required in $V_n$. The result in Section~3 shows that
under the null hypothesis $H_0$, $\hat q\rightarrow 1$, which can avoid the curse of dimensionality greatly. Another superiority of the new test is the model-adaptive property, that is, through estimating the matrix $B$, the test can automatically adapt the  hypothetical and alternative model such that it can have better performance in the significance level maintainance and power enhancement. To be specific,  under $H_0$, $\hat B(\hat q)\rightarrow c\beta$ for a constant $c$, and under $H_1$ $\hat q\rightarrow q\geq 1$, $\hat B(\hat q)\rightarrow BC$ for a $q\times q$ orthogonal matrix, adaptive to the alternative model (\ref{alter_model}).
\end{remark}

\begin{remark}
Another related test is the dimension reduction model-adaptive test $T_n^{GWZ}$ proposed by Guo et al. (2015):
\begin{equation}\label{T_GWZ}
    T_n^{GWZ}=\frac{1}{n(n-1)} \sum_{i=1}^n\sum_{j\neq i}^n \mathcal{K}_h\{\hat B(\hat q)^\top (x_i-x_j)\}\hat e_i\hat e_j.
\end{equation}
In the Section~\ref{sec4}, we will show through Hampel influence function analysis that RDREAM has more stable and robust behavior than $T_n^{GWZ}$ when the response is contaminated.
\end{remark}

\subsection{Identification and estimation of $B$}
As the estimates of the matrix $B$ and structural dimension $q$ are crucial for our RDREAM, we first
specify the estimate for the matrix $B$ under given $q$ and then study how to select $q$ consistently. Note that the model (\ref{alter_model}) is a multi-index model with unknown $q$ indexes, thus outer product of gradients (OPG) introduced by Xia et al (2002) can be considered to estimate $B$. Another method to estimate $B$ is inspired by sufficient dimension reduction technique. In fact, $B$ is not identifiable since for any $q\times q$ orthogonal matrix $C$, $m(B^\top X)$ can always be rewritten as $\tilde m(C^\top B^\top X)$. Therefore, what we can identify is the space spanned by $B$ via sufficient dimension reduction technique, or in other words, we can identify $q$ base vectors of  the space spanned by $B$. There exist several proposals in the literature, such as sliced inverse regression (SIR) proposed by Li (1991), sliced average variance estimation (SAVE) considered by Cook and Weisberg (1991), minimum average variance estimation (MAVE) advised by Xia et al (2002) and discretization-expectation estimation (DEE) suggested by Zhu et al (2010). In view of easy-operation and good-performance of DEE, we consider to employ it to estimate $B$. Further, as the method called  outer product of the gradients in Xia et al (2002) has less restriction on the covariates $X$, we then also use it to estimate $B$ for a comparison with SIR-based DEE. In the following, we give simple review of OPG and SIR-based DEE.


\subsubsection{Outer product gradients}
The outer product of the gradients can be written as
\begin{equation*}
    E\{\nabla m(B^\top X)\nabla m(B^\top X)^\top\}=BE\{m'(B^\top X)m'(B^\top X)^\top\}B^\top,
\end{equation*}
where
$m(B^\top x)=E(Y|X=x)$, $\nabla m(B^\top X)=\frac{\partial}{\partial X}m(B^\top X)$ and
$m'(B^\top X)=\frac{\partial}{\partial B^\top X}m(B^\top X)$. Note that $E\{\nabla m(B^\top X)\nabla m(B^\top X)^\top\}$ has $q$ nonzero eigenvalues, the matrix $B$ is in the space spanned by the $q$ eigenvectors corresponding to the largest $q$ eigenvalues of $E\{\nabla m(B^\top X)\nabla m(B^\top X)^\top\}$. Thus, we need to estimate the expectation $E\{\nabla m(B^\top X)\nabla m(B^\top X)^\top\}$ and then obtain the estimator of $B$.

By local linear fitting
\begin{equation*}
    m(B^\top x_i)=m(B^\top x_j)+m'(B^\top x_j)^\top B^\top (x_i-x_j)=a_j+b_j^\top x_{ij},
\end{equation*}
where $a_j=m(B^\top x_j)$, $b_j=B\times m'(B^\top x_j)$ and $x_{ij}=x_i-x_j$, we can obtain $(
\hat a_j,\hat b_j)$ by minimizing the following objective function
\begin{equation}\label{OPG_ajbj}
    \min_{a_j,b_j}\sum_{i=1}^n \mathcal{K}_h(B^\top x_{ij})\{y_i-a_j-b_j^\top x_{ij}\}^2,
\end{equation}
where $\mathcal{K}_h(\cdot)=\mathcal{K}(\cdot/h)/h$ with $\mathcal{K}(\cdot)$ being a $q$-dimensional kernel function and $h$ being a bandwidth. The corresponding estimating equation from (\ref{OPG_ajbj}) for $(a_j,b_j)$ can be given as
\begin{equation*}
    \sum_{i=1}^n \mathcal{K}_h(B^\top x_{ij})(1,x_{ij}^\top)^\top\{y_i-\hat a_j-\hat b_j^\top x_{ij}\}=0.
\end{equation*}
Further, $E\{\nabla m(B^\top X)\nabla m(B^\top X)^\top\}$ can be estimated as
\begin{equation}\label{hatSigma_OPG}
    \hat\Sigma=\frac{1}{n}\sum_{j=1}^n\hat b_j\hat b_j^\top.
\end{equation}
Therefore, the $q$ eigenvectors corresponding to the largest $q$ eigenvalues of $\hat\Sigma$ can be regarded as the estimator of the matrix $B$.

\subsubsection{Discretization-expectation estimation}
We first give some notations. The central subspace, denoted by $\mathcal{S}_{Y|X}$, is defined as the intersection of all subspaces  spanned by the columns of a matrix $A$, $\mathcal{S}_A$, of minimal dimension such that $Y\hDash X|A^\top X$, where $\hDash$ stands for statistical independence. Similarly,
let $\mathcal{S}_{E(Y|X)}$ be the intersection of all subspaces $\mathcal{S}_A$ spanned by the matrix $A$ such that $Y\hDash E(Y|X)|A^\top X$. In sufficient dimension reduction, $\mathcal{S}_{E(Y|X)}$ is called the central mean subspace and its dimension, denoted by $d_{E(Y|X)}$, is called the structural dimension. As to the model (\ref{alter_model}), we have $\mathcal{S}_{E(Y|X)}=span(B)$ and $d_{E(Y|X)}=q$. Thus, we aim to identify the $q$ base vectors of $\mathcal{S}_{E(Y|X)}$.

Compared with SIR, DEE can avoid the choice of the number of slices,  as Li and Zhu (2007) pointed out, which may affect the efficiency and even lead to inconsistent estimates, and has no optimal solution. Define the new response variable $Z(t)=I(Y\leq t)$ for any $t$, where the indicator function $I(Y\leq t)$ takes the value $1$ if $Y\leq t$ and $0$, otherwise. When SIR is applied, the original related matrix $\mathcal{M}(t)$ based on SIR is a $p\times p$ positive semi-definite matrix such that $span\{\mathcal{M}(t)\}=\mathcal{S}_{Z(t)|X}$. Here, $\mathcal{S}_{Z(t)|X}$ is the central subspace of $Z(t)|X$. Given $\mathcal{M}=E\{\mathcal{M}(T)\}$, according to Theorem~1 in Zhu et al (2010), $span\{\mathcal{M}\}=\mathcal{S}_{Y|X}$. To ensure
$\mathcal{S}_{Y|X}=\mathcal{S}_{E(Y|X)}$ for $\varepsilon$ in the model (\ref{alter_model}), based on Guo et al (2015), a condition that $\varepsilon=m_1(B^\top X)\tilde{\varepsilon}$ and $\tilde{\varepsilon}\hDash X$ is needed.

Based on the above analysis, estimating $\mathcal{S}_{E(Y|X)}$ amounts to estimating $\mathcal{M}$. Given the sample $\{(y_1,x_1), (y_2,x_2),\cdots,(y_n,x_n)\}$, we define the dichotomized responses as $z_i(y_j)=I(y_i\leq y_j),\,\,i,j=1,\ldots,n$. Thus, for each fixed $y_j$, we can obtain a new sample $\{(z_1(y_j),x_1), (z_2(y_j),x_2),\cdots,(z_n(y_j),x_n)\}$ and the estimate $\mathcal{M}_n(y_j)$ of $\mathcal{M}(y_j)$ can be gained with SIR. Thus,  $\mathcal{M}$ can be estimated as
\begin{equation}\label{hatM_DEE}
    \mathcal{M}_{n,n}=n^{-1}\sum_{j=1}^n\mathcal{M}_n(y_j),
\end{equation}
which has been proved  to be  root-$n$ consistent to  $\mathcal{M}$ by Zhu et al (2010). The $q$ eigenvectors of $\mathcal{M}_{n,n}$ corresponding to its $q$ largest eigenvalues are applied to estimate $B$. For this method, a mild linearity condition is assumed: $E(X|B^\top X=u)$ is linear in $u$ (Li, 1991).

\subsection{Estimation of structural dimension $q$}
In order to get  RDREAM in (\ref{statistic_Vn}), the estimate of structural dimension $q$ is necessary for the above two methods of identifying $B$. Here, a Ridge-type Ratio Estimate (RRE) method, which is inspired by Xia et al (2015), is suggested to determine $q$ for OPG and DEE. It is based on the ratios of the eigenvalues  with an artificially added ridge value $c$. Denote $\hat \lambda_1\geq \hat \lambda_2\geq \ldots\geq \hat \lambda_p$ to be the eigenvalues of the estimating matrix $\hat\Sigma$ or $\mathcal{M}_{n,n}$ in (\ref{hatSigma_OPG}) and (\ref{hatM_DEE}), respectively. $\hat q$ can be obtained as
\begin{equation*}
    \hat q=\arg\min_{k=1,\ldots,p-1}\frac{\lambda_{k+1}+c}{\lambda_k+c},
\end{equation*}
where the constant $c=1/\sqrt{nh}$ is recommended. The consistencies of $\hat q$ under the null hypothesis (\ref{null_model}) and global alternative hypothesis (\ref{alter_model}) are shown in the following lemma.
\begin{lemma}\label{hatq_consist}
Assume that  the OPG-based matrix $\hat \Sigma$ or  the DEE-based matrix $M_{nn}$ is root-$n$ consistent to $\Sigma$ or $M$. Then the corresponding estimate $\hat q= q$ as $n\rightarrow\infty$ with a probability going to one. Therefore, for a $q\times q$ orthogonal matrix $C$, $\hat B(\hat q)$ is a root-$n$ consistent estimate of $BC^\top$.
\end{lemma}


\section{Asymptotic properties}\label{sec3}
\renewcommand{\theequation}{3.\arabic{equation}}
\setcounter{equation}{0}

In this section, the large-sample properties of the  RDREAM test statistic $V_n$ in (\ref{statistic_Vn})
are investigated via its asymptotic distributions under the null hypothesis, global alternative hypothesis and local alternative hypothesis.

\subsection{Limiting null distribution}
The asymptotic normality discussed in the following also requires  that the regression parameter is root-$n$ consistently estimated under $H_0$ and the residuals must come from a robust fit.
Let $Z=B^\top X$ and
\begin{equation*}
    Var=\frac{1}{72}\int \mathcal{K}^2(u)du\int p^2(z)dz,
\end{equation*}
where $p(z)$ is the probability density function of $Z$. Moreover, $Var$ can be consistently estimated by:
\begin{equation*}
    \widehat{Var}=\frac{1}{72n(n-1)}\sum_{i=1}^n\sum_{j\neq i}^n \frac{1}{h^{\hat q}}\mathcal{K}^2\Big\{\frac{\hat B(\hat q)^\top (x_i-x_j)}{h}\Big\}.
\end{equation*}

We first state the asymptotic property of  the RDREAM test statistic  in (\ref{statistic_Vn}) under the null hypothesis $H_0$ as follows:

\begin{theorem}\label{theo_null}
Suppose that conditions (C1)-(C8) in the Appendix hold. Under $H_0$, we have
\begin{equation*}
    nh^{1/2}V_n\Rightarrow N(0, Var).
\end{equation*}
\end{theorem}

Plugging in a consistent estimator of $Var$, a standardized version of  $V_n$ can be defined as
\begin{equation}\label{S_n}
    S_n=\frac{n-1}{n}\frac{nh^{1/2}V_n}{\sqrt{\widehat{Var}}}.
\end{equation}
The following corollary can be easily obtained.
\begin{corollary}\label{coro1}
Under  $H_0$ and  Conditions (C1)-(C8) in the Appendix, we have
\begin{equation*}
    S_n^2\Rightarrow \chi_1^2,
\end{equation*}
where $\chi_1^2$ is the chi-square distribution with one degree of freedom.
\end{corollary}

Theorem~\ref{theo_null} and Corollary~\ref{coro1} characterize the asymptotic properties of the test statistic $V_n$. Based on Corollary~\ref{coro1}, $p$-values of RDREAM can be easily determined by the quantiles of the chi-square distribution with one degree of freedom. The null hypothesis $H_0$ is rejected when $S_n\geq \chi^2_{1-\alpha}(1)$ where $\chi^2_{1-\alpha}(1)$ is the $1-\alpha$ upper quantile of the chi-square distribution.


\subsection{Power study}
We are now in the position to examine the power performance of our RDREAM under alternative hypothesis. More specifically, the following sequence of  alternative models is  under consideration:
\begin{equation}\label{loca_alter}
    H_{1n}: \quad Y=g(\tilde{\beta}^\top X,\tilde{\theta})+C_n m(B^\top X)+\varepsilon,
\end{equation}
where $E(\varepsilon|X)=0$, $E[m^2(B^\top X)]<\infty$ and $\{C_n\}$ is a constant sequence. When $C_n=C$ for a nonzero constant $C$, the model is a global alternative model, while when $C_n$ goes to zero, it is a sequence of local alternative models. In this sequence of models, $\beta$ is one of the column in $B$.
Denote $\tilde\alpha=(\beta,\theta)$ to be
\begin{equation*}
    \tilde\alpha=\arg\min_\alpha E\{g(\beta^\top X,\theta)-m(X)\},
\end{equation*}
where $m(X)=E(Y|X)$. When the null hypothesis $H_0$ holds, $\tilde\alpha$ is the true parameter. For a robust estimate $\hat\alpha$, we have $\hat\alpha-\tilde\alpha=O_p(1/\sqrt{n})$. We first discuss the consistency of $\hat q$ under the local alternative hypothesis (\ref{loca_alter}). When $n\rightarrow\infty$, the local alternative models converge to the null model, $\hat q$s under the local alternative models are expected to converge to $\hat q$ under the null model, which finally converge to the structural dimension $q=1$ under the null model.
\begin{lemma}\label{consis_qlocal}
Assume  conditions (C1)-(C8) in the Appendix hold and under the local alternative hypothesis (\ref{loca_alter}) with $C_n=n^{-1/2}h^{-1/4}$, we have $\hat q=1$ as $n\rightarrow\infty$ with a probability going to one, where $\hat q$ is either the  OPG-based  or the DEE-based estimate.
\end{lemma}

The asymptotic properties under global and local alternative hypotheses are concluded in the following Theorem.
\begin{theorem}\label{theo_alter}
Under Conditions (C1)-(C8) in the Appendix, we have:

(i) Under the global alternative of (\ref{alter_model}) or equivalently the above model with $C_n=C$,
\begin{equation*}
    S_n/(nh^{1/2})\Rightarrow Constant>0;
\end{equation*}

(ii) Under the local alternative hypothesis (\ref{loca_alter}) with $C_n=n^{-1/2}h^{-1/4}$, we have
\begin{equation*}
    nh^{1/2}V_n\Rightarrow N(\mu, Var)~~~\mbox{and}~~~S_n^2\Rightarrow \chi_1^2(\mu^2/Var),
\end{equation*}
where
\begin{equation*}
    \mu=E[h^2(\varepsilon)m^2(B^\top X)p(X)],
\end{equation*}
$h(\cdot)$ denotes the probability density function of $\varepsilon$
and $\chi_1^2(\mu^2/Var)$ is a noncentral chi-squared random variable with one degree of freedom and the non-centrality parameter $\mu^2/Var$.
\end{theorem}

The above Theorem indicates that under the global alternative hypothesis, the RDREAM test is
consistent with the asymptotic power $1$ and  can detect the local alternatives distinct from the null hypothesis at a nonparametric rate of order $n^{-1/2}h^{-1/4}$, which is the optimal rate with one dimensional predictor for the test $\tilde V_n$ in (\ref{statistic_Vnwan}).

\begin{remark}
From the above theorem, we can observe that under the null hypothesis, the use (automatically) of lower order kernel function in  $V_n$ in (\ref{statistic_Vn}) makes a very significant improvement than $\tilde V_n$ in (\ref{statistic_Vnwan}). Based on Theorem~\ref{theo_null}, $V_n$ owns a much faster convergence rate of order $nh^{1/2}$ and $nh^{1/2}V_n$ is asymptotically  normal under the null whereas the rate of order $nh^{p/2}$ is for $\tilde V_n$. Further, according to Theorem~\ref{theo_alter}, the conclusion can be made that $V_n$ is much more sensitive than Wang and Qu's test $\tilde V_n$ in the sense that $V_n$ can detect the local alternatives distinct from the null at the rate of $n^{-1/2}h^{-1/4}$ whereas $\tilde V_n$ is only workable at the rate of order $n^{-1/2}h^{-p/4}$. Therefore, the power performance of the proposed test can be much enhanced.
\end{remark}

\begin{remark}
Our original simulation results based on $S_n$ in (\ref{S_n}) suggest the conservative sizes of tests. Thus, the following size-adjustment is needed for the test statistics with both the OPG-based estimate and the DEE-based estimate:
\begin{equation}\label{adjusted_formula}
    \tilde{S}_n=(1+4n^{-4/5})S_n.
\end{equation}
The size-adjustment constant is chosen through intensive simulation with various different values and this one is recommended. With such a size-adjustment, our new test $\tilde{S}_n$ can better control type I errors. It is worth noting that the size-adjustment is asymptotically negligible when $n\rightarrow\infty$ since $\tilde{S}_n\rightarrow S_n$ when $n\rightarrow\infty$.
\end{remark}

\section{Robustness property}\label{sec4}
\renewcommand{\theequation}{4.\arabic{equation}}
\setcounter{equation}{0}

In this section, we investigate the local stability of  RDREAM under infinitesimal local contamination through Hmapel influence function, which was introduced by Hampel (1974). The Hampel influence analysis  reveals that RDREAM has the desired local robustness property. In the following, we first give von Mises functional expansion of RDREAM and further derive the Hampel influence function. For this investigation, Wang and Qu (2007) is a good reference. The key difference is between the used covariate $X$ and $\hat B(\hat q)^\top X$ in the respective test statistics, and $\hat B(\hat q)^TX$ is automatically $1$- and $q$-dimensional under the null and alternative hypothesis. Thus, we only give some brief descriptions about the results.

Following Wang and Qu (2007), the von Mises analysis (see Fernholz (1983)) can provide the basis of  the Hampel influence function calculation. We now discuss the von Mises functional expansion.

Let $ Z= B^\top X$, $\hat Z=\hat B(\hat q)^\top X$ and denote $\hat H_n$ as the empirical distribution function of $Y_1-g(\hat\beta^\top X_1,\hat\theta),
\ldots,Y_n-g(\hat\beta^\top X_n,\hat\theta)$. When $H_0$ holds, $B=c\beta$, and $\hat B(\hat q)$ is an estimator of $\beta$  up  to a scalar constant $c$.  The RDREAM statistic $V_n$ in (\ref{statistic_Vn}) can be asymptotically expressed as
\begin{eqnarray*}
  T(\hat F_h,F_n)&=:&\int\int\Big[\hat H_n(y-g((\beta^\top x)(F_n),\theta(F_n)))-\frac{1}{2}\Big]\times\Big(
  \int\Big[\hat H_n(y_1-g((\beta^\top x)(F_n),\theta(F_n)))\\
  &&-\frac{1}{2}\Big]\times\hat f_h(\hat z,y_1)dy_1\Big)d F_n(\hat z,y),
\end{eqnarray*}
where $F_n(\cdot)$ is the empirical distribution function of $(\hat Z_i, Y_i)$'s, $((\beta^\top x)(F_n),\theta(F_n))^\top$ is an estimator of $(\beta^\top x,\theta)^\top$, which can be rewritten as a functional of the empirical distribution $F_n$, and $\hat f_h(z_i,y_i)$ is a smoothing kernel estimation of the joint density function of $(\hat Z,Y)$, which has the following form:
\begin{equation*}
    \hat f_h(\hat z_i,y_i)=\frac{1}{n-1}\sum_{i\neq j}^n \mathcal{K}_{1,h}(y_i-y_j)
    \mathcal{K}_{2,h}(\hat z_i-\hat z_j),
\end{equation*}
where $\mathcal{K}_{1}(\cdot),\mathcal{K}_{2}(\cdot)$ are two kernel functions and satisfy $\mathcal{K}_{1,h}(\cdot)=\mathcal{K}_{1}(\cdot/h)/h$ and $\mathcal{K}_{2,h}(\cdot)=\mathcal{K}_{2}(\cdot/h)/h^{\hat q}$, respectively.
An appropriate functional for  RDREAM is  bivariate with the form
\begin{eqnarray}
  &T(F,F)&\nonumber\\
  &=:&\int\int\Big[H(y-g((\beta^\top x)(F),\theta(F)))-\frac{1}{2}\Big]\times\Big(
  \int\Big[H(y_1-g((\beta^\top x)(F),\theta(F)))-\frac{1}{2}\Big]\nonumber\\
  &&\times f(z,y_1)dy_1\Big)d F(z,y),\label{T_F}
\end{eqnarray}
here, $F(\cdot)$ is the distribution function of $(Z,Y)$ and $H(\cdot)$ is the distribution function of $Y-g(\beta^\top X,\theta)$: $H(v)=\int\int_{y\leq g((\beta^\top x)(F),\theta(F))+v}dF(z,y)$.

\

Now we are in the position to analyse Hampel influence function (Hampel 1974). When the second-order influence function in the response direction,  the regression functions is robustly estimated. Thus, we calculate the first-order and second-order influence function of $T(F,F)$ in (\ref{T_F}) at the point $(z_0,y_0)$.
When  $H_0$ holds, the Hampel's first-order influence function of $T(F,F)$ in (\ref{T_F}) at the point $(z_0,y_0)$ is defined as
\begin{equation*}
    IF^{(1)}(z_0,y_0;T)=\lim_{t\rightarrow 0}\frac{T(F_t,F_t)-T(F,F)}{t},
\end{equation*}
where $T(F,F)=0$ and $F_t=(1-t)F+t\Delta_{z_0,y_0}$, here, $\Delta_{z_0,y_0}$ is the point mass function at the point $(z_0,y_0)$. Denote $L(t)=T(F_t,F_t)=T(F+tU,F+tU)$, where $U=\Delta_{z_0,y_0}-F$. Thus, we have $L(0)=0$. From the proof in the Appendix, it is not difficult to obtain that $\frac{d L(t)}{dt}|_{t=0}=0$. Therefore,  RDREAM has  a degenerate first-order influence function.

To obtain the second-order influence function of RDREAM, we first compute
\begin{eqnarray}\label{sec_deri}
  && \frac{1}{2}\frac{d^2}{dt^2}L(t)|_{t=0}\nonumber\\
  &=&\int\Big(\int\dot{H}(y-g((\beta^\top x)(F),\theta(F)))f(z,y)dy\Big)^2dz\nonumber\\
&&+\int\int\dot{H}(y-g((\beta^\top x)(F),\theta(F)))\times\int \Big[H(y_1-g((\beta^\top x)(F),\theta(F)))-\frac{1}{2}\Big]\nonumber\\
&&\times u(z,y_1)dy_1dF(z,y)+\int\int\Big[H(y-g(\beta^\top x(F),\theta(F)))-\frac{1}{2}\Big]\nonumber\\
&&\times\int\dot{H}(y_1-g((\beta^\top x)(F),\theta(F)))\times f(z,y_1)dy_1dU(z,y)\nonumber\\
&&+\int\int\Big[H(y-g((\beta^\top x)(F),\theta(F)))-\frac{1}{2}\Big]\times\Big(\int\Big[H(y_1-g((\beta^\top x)(F),\theta(F)))\nonumber\\
&&-\frac{1}{2}\Big]u(z,y_1)dy_1\Big)d U(z,y),
\end{eqnarray}
where $\dot{H}(\cdot)=:\frac{d H_t(\cdot)}{dt}|_{t=0}$ and $H_t(\cdot)$ represents $H(\cdot)$ under contamination, that is, $H_t(v)=\int\int_{y\leq g(\beta^\top x(F_t),\theta(F_t))+v}dF_t(z,y)$.  Besides, $u(z,y)$ is the probability density function of $U(z,y)$. Taking $U=\Delta_{z_0,y_0}-F$ into the formula (\ref{sec_deri}), it is shown that the four terms in (\ref{sec_deri}) converge at the same rate. Further, based on Hampel's definition, we can obtain the second-order influence function of RDREAM at the point $(z_0,y_0)$ as follows:
\begin{eqnarray*}
  IF^{(2)}(z_0,y_0)&=&\int\Big(\int\dot{H}_{\Delta_{(z_0,y_0)}}(y-g((\beta^\top x)(F),\theta(F)))f(z,y)dy\Big)^2dz\nonumber\\
  &&+\Big[H(y_0-g((\beta^\top x_0)(F),\theta(F)))-\frac{1}{2}\Big]\times \nonumber\\
  &&\int\int\dot{H}_{\Delta_{(z_0,y_0)}}(y-g((\beta^\top x)(F),\theta(F)))d F(z,y)\nonumber\\
  &&+\Big[H(y_0-g((\beta^\top x)_0(F),\theta(F)))-\frac{1}{2}\Big]\times\int\dot{H}_{\Delta_{(z_0,y_0)}}(y-g(\beta^\top x_0(F),\theta(F)))\nonumber\\
  &&\times f(z_0,y)dy+\Big[H(y_0-g((\beta^\top x)_0(F),\theta(F)))-\frac{1}{2}\Big]^2,
\end{eqnarray*}\label{IF_RDREAM}
where $z_0=\beta^\top x_0$ and $\dot{H}_{\Delta_{(z_0,y_0)}}(y-g((\beta^\top x)(F),\theta(F)))$ denotes $\frac{d}{dt}H(y-g((\beta^\top x)(F_t),\theta(F_t)))|_{t=0,U=\Delta_{(z_0,y_0)}-F}$.
The detail expression of $\dot{H}_{\Delta_{(z_0,y_0)}}(y-g((\beta^\top x)(F),\theta(F)))$ can be written as
\begin{eqnarray*}
  &&\dot{H}_{\Delta_{(x_0,y_0)}}(y-g((\beta^\top x)(F),\theta(F)))\\
  &=&\int h(y-g((\beta^\top x)(F),\theta(F)))\times\mbox{grad}_{\alpha}\{g((\beta^\top x)(F),\theta(F))\}^\top\times\frac{d\alpha}{dt}d F_X(x)\\
  &&+I\Big(y_0\leq y+g((\beta^\top x)_0(F),\theta(F))-g((\beta^\top x)(F),\theta(F))\Big)\\
  &&-H(y-g((\beta^\top x)(F),\theta(F))),
\end{eqnarray*}
where $\alpha=(\beta,\theta)^\top$ and $\mbox{grad}_{\alpha}\{g((\beta^\top x)x(F),\theta(F))\}^\top$ represents the gradient of $g((\beta^\top x)(F),\theta(F))$ with respect to $\alpha$. When the parameter $\alpha$ comes from a robust fit, we have that $d\alpha/dt$ is bounded. Together with the conditions (C1) and (C5) in the Appendix, it can be shown that $\dot{H}_{\Delta_{(z_0,y_0)}}(y-g((\beta^\top x)(F),\theta(F)))$ is also bounded.
Further, the second-order influence function $IF^{(2)}(z_0,y_0)$ of RDREAM is bounded in the response direction.

For the purpose of comparison, we can similarly derive the first-order and second-order influence function for the test $T^{GWZ}_n$ in (\ref{T_GWZ}). the first-order  influence function is also zero and the second-order influence function can be derived as
\begin{eqnarray*}
  IF_{GWZ}^{(2)}(z_0,y_0)&=&\int\Big(\frac{d}{dt}g((\beta^\top x)(F),\theta(F))f(x)\Big)^2dz+[y_0-g((\beta^\top x)_0(F),\theta(F))]^2\nonumber\\
  &&-[y_0-g((\beta^\top x)_0(F),\theta(F))]\int\frac{d}{dt}g((\beta^\top x)(F),\theta(F))dz\nonumber\\
  &&-[y_0-g((\beta^\top x)_0(F),\theta(F))]\frac{d}{dt}g((\beta^\top x)_0(F),\theta(F))f(z_0),
\end{eqnarray*}\label{IF_GWZ}
where $\frac{d}{dt}g((\beta^\top x)(F),\theta(F))=\frac{d}{dt}g((\beta^\top x)(F_t),\theta(F_t))|_{t=0,U=\Delta_{(z_0,y_0)}-F}$.

The second-order influence function $IF_{GWZ}^{(2)}(z_0,y_0)$ in the $y$-direction is not bounded.

The above influence function analysis indicates RDREAM possesses more stable and robust performance than the test $T_n^{GWZ}$ when the response is under contamination.

\section{Simulation studies}\label{sec5}
\renewcommand{\theequation}{5.\arabic{equation}}
\setcounter{equation}{0}

In this section, three simulation studies are conducted to examine the theory and the finite-sample performance of the proposed RDREAM. Throughout this section, denote the adjusted RDREAM statistic in the formula (\ref{adjusted_formula}) based on the OPG and SIR-based DEE estimate as $\tilde{S}_n^{OPG}$ and $\tilde{S}_n^{DEE}$, respectively. The purpose of the simulation studies is three-fold: to examine the robustness of the new method; to check the usefulness to overcome the curse of  dimensionality; to demonstrate its usefulness in the cases without outliers.  To this end,  the objective of the first study is to check and compare the performance of RDREAM: $\tilde{S}_n^{OPG}$ and $\tilde{S}_n^{DEE}$. The effects of different distributions of the error and nonlinearity under the null hypothesis on the performance of the new tests is also considered in this study. The second study is used to examine the impact from dimensionality on both RDREAM and the robust test $T_n^{WQ}$ introduced by Wang and Qu (2007). Since the test $T_n^{GWZ}$ proposed by Guo et al (2015) is also to solve the dimensionality problem in model checking, the third study aims to show the robustness properties of the proposed test via comparing $\tilde{S}_n^{OPG}, \tilde{S}_n^{DEE}$ with $T_n^{GWZ}$ and to examine its performance in the case without outliers.

\vspace{0.2cm}
{\it Study~1}: Consider the following models
\begin{eqnarray*}
  &&H_{11}: Y=\beta^\top X+a\times \exp(-1.5\beta^\top X)+\varepsilon,\\
  &&H_{12}: Y=\beta^\top X+1.5a(\beta^\top X)^3+\varepsilon,\\
  &&H_{13}: Y=\beta^\top X+6a\times \cos(0.8\pi\beta^\top X)+\varepsilon,\\
  &&H_{14}: Y=2.5\exp(0.5\beta^\top X)+1.5a(\beta^\top X)^3+\varepsilon,
\end{eqnarray*}
where $p=8$, $\beta=(1,\ldots,1)^\top/\sqrt{p}$. The covariate $X=(X_1,\ldots,X_p)^\top$ are i.i.d and generated from a multivariate normal distribution $N(0,I_p)$ where $I_p$ is a $p\times p$ identity matrix. Consider two kinds of errors: one is $\varepsilon\sim N(0,1)$ and the other is that $\varepsilon_i$'s are i.i.d from a log-normal distribution $\ln N(0,0.25)$ that is standardized to have mean $0$ and variance $1$. For all of models, $10\%$ of the responses are randomly added by an outlying value $5$. We set $a=0,0.2,\ldots,1.0$ where $a=0$ corresponds to the null hypothesis and $a\neq 0$ corresponds to the alternative hypothesis. For $H_{11},H_{12}$ and $H_{13}$, the null models are a linear model and the alternative models are all single-index models. The null model of $H_{14}$ is a nonlinear model. Under the alternatives, the third model $H_{13}$ is high-frequent and the other three models are not. We intend to examine whether the new tests can be powerful for both the two types of  models.

To compute  the robust test statistics $\tilde{S}_n^{OPG}$ and $\tilde{S}_n^{DEE}$, the residuals are obtained from a robust regression through $M$-estimate. As to the nonparametric regression estimation, throughout these simulations, unless otherwise specified, the kernel function is taken to be $\mathcal{K}(u)=15/16(1-u^2)^2$ if $|u|\leq 1$ and $0$, otherwise. Our experience in the simualtions suggests that RDREAM is not sensitive to the choice of kernel function.
The bandwidth is recommended as $h_{OPG}=1.8n^{-1/(\hat q+4)}$ for the test statistic $\tilde{S}_n^{OPG}$ and $h_{DEE}=0.5n^{-1/(\hat q+4)}$ for $\tilde{S}_n^{DEE}$ through intensive numerical computation. The significance level is set to be $\alpha=0.05$ and the sample size $n=60,100,200$ are considered. Every simulation result is the average of $2000$ replications.

Table~\ref{tab_study1} displayed the empirical sizes and powers of the new  tests against the alternatives $H_{11},H_{12}$ with different values of $a$. From this table, we can see that for every case we conduct, the adjusted test statistics $\tilde{S}_n^{OPG}$ and $\tilde{S}_n^{DEE}$ can maintain the significance level very well even with the sample size $n=60$. It is reasonable that the larger the sample size is, the closer the empirical size is to the significance  level. As to the  power performance, for any a specific test $\tilde{S}_n^{OPG}$ or $\tilde{S}_n^{DEE}$, the  powers  are higher with larger sample sizes. $\tilde{S}_n^{OPG}$ and $\tilde{S}_n^{DEE}$, from the results of $H_{11}$, it can be seen that in most cases, $\tilde{S}_n^{OPG}$ outperforms  $\tilde{S}_n^{DEE}$; for $H_{12}$, $\tilde{S}_n^{DEE}$ exhibits slightly higher powers than  $\tilde{S}_n^{OPG}$, however, the difference between them can be negligible. Therefore, the test $\tilde{S}_n^{OPG}$ yields to no worse  power performance than $\tilde{S}_n^{DEE}$. At last, for every specific alternative model, the impact of error distribution on empirical sizes and  powers shows no significant difference, which, to some extent, illustrates that the proposed tests are robust to light and heavy tail error distributions.
\begin{center}
Table~\ref{tab_study1} about here
\end{center}

\begin{figure}[htbp]
\centering
\includegraphics[width=14cm,height=12cm]{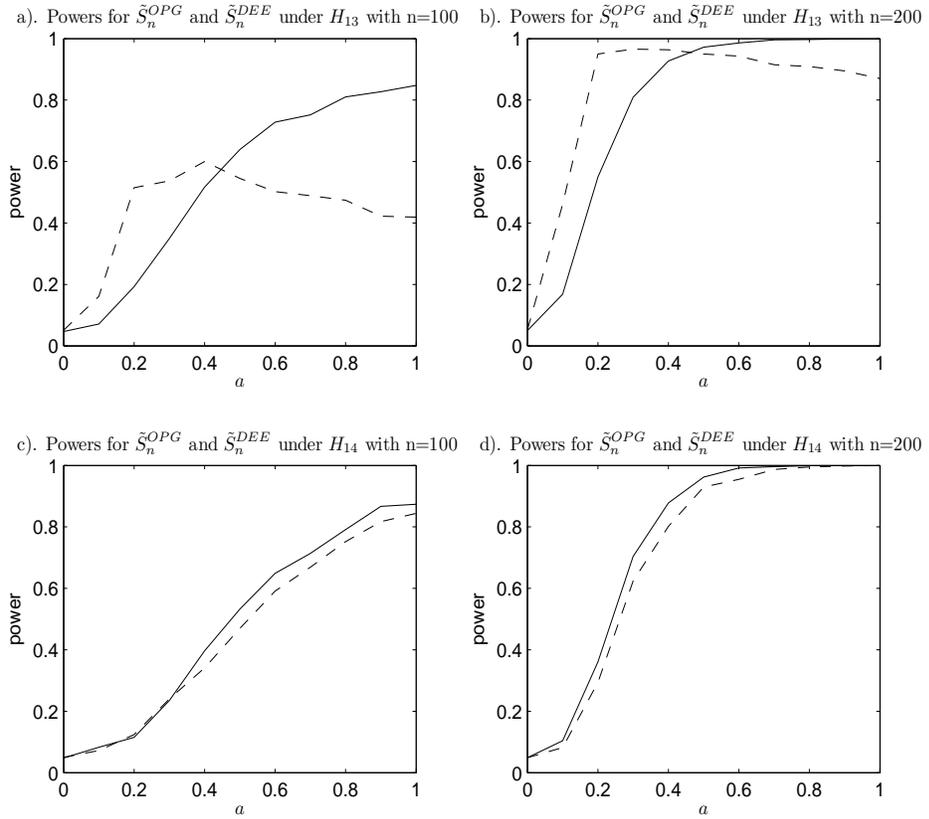}
 \caption{\small
 Empirical sizes and powers of $\tilde{S}_n^{OPG}$ and $\tilde{S}_n^{DEE}$ for $H_{10}$ v.s. $H_{13},H_{14}$ at the significance level $\alpha=0.05$ with $p=8$, $X\sim N(0,I_p)$ and $\varepsilon\sim \ln N(0,0.25^2)$.  In four plots, the solid line and the dash line are for $\tilde{S}_n^{OPG}$ and $\tilde{S}_n^{DEE}$, respectively.
  }
  \label{plotH1314}
\end{figure}

To get a sense of the performance under high-frequent alternative model and the effect of nonlinear null model on the proposed tests, we conduct a more in-depth analysis for the alternatives $H_{13}$ and $H_{14}$. The simulated power curves of different values of $a$ and sample sizes $n=100,200$ are displayed in Figure~\ref{plotH1314}. From this figure,  we can see that for the cosine alternative $H_{13}$, the simulated power curve for $\tilde{S}_n^{OPG}$ has a sigmoidal shape, whereas the curve for $\tilde{S}_n^{DEE}$ shows rapid growth at the beginning and slightly decrease for a relative larger $a$. The different power performance for $\tilde{S}_n^{OPG}$ and $\tilde{S}_n^{DEE}$ may be ascribed to the different sensitivity to high-frequent alternative model. However, their powers  are all acceptable. As to the alternative $H_{14}$, both of these two tests show the popular sigmoidal shape power curves and $\tilde{S}_n^{OPG}$ slightly outperforms $\tilde{S}_n^{DEE}$.
\begin{center}
Figure~\ref{plotH1314} about here
\end{center}

In summary, the proposed tests $\tilde{S}_n^{OPG}$ and $\tilde{S}_n^{DEE}$ can both control type I error very well and make an excellent power performance. The OPG-based test $\tilde{S}_n^{OPG}$ is more powerful.

\vspace{0.2cm}
{\it Study~2}: The data are generated from the following models:
\begin{eqnarray*}
   &&H_{21}: Y=\beta_1^\top X+1.5a(\beta_2^\top X)^3+\varepsilon,\\
   &&H_{22}: Y=\beta_1^\top X+0.3a\Big\{4(\beta_2^\top X)^3+(\beta_2^\top X)^2\Big\}+\varepsilon,\\
   &&H_{23}: Y=\beta_1^\top X+4a\exp(-\beta_2^\top X)+\varepsilon,
\end{eqnarray*}
where $\beta_1=(1,\ldots,1)^\top/\sqrt{p}$, $\beta_2=(\underbrace{1,\ldots,1}_{p/2},0,\ldots,0)^\top/\sqrt{p/2}$, $p=4,2$ and $n=100,200$.
When $a\neq 0$, we have $q=2$ and $B=(\beta_1,\beta_2)$. The covariates $X$ and the  error $\varepsilon$ is independently generated from the multivariate and univariate standard normal distributions, respectively. For all of cases, $10\%$ of the responses are randomly replaced by observations from a nonlinear model $Y=5.5\cos(3\pi\beta_1^\top X)+\varepsilon$. We intend to apply these alternative models to examine the effect of dimensionality on the proposed tests $\tilde{S}_n^{OPG},\tilde{S}_n^{DEE}$ and $T_n^{WQ}$ considered by Wang and Qu (2007).

\begin{figure}[htbp]
\centering
\includegraphics[width=14cm,height=12cm]{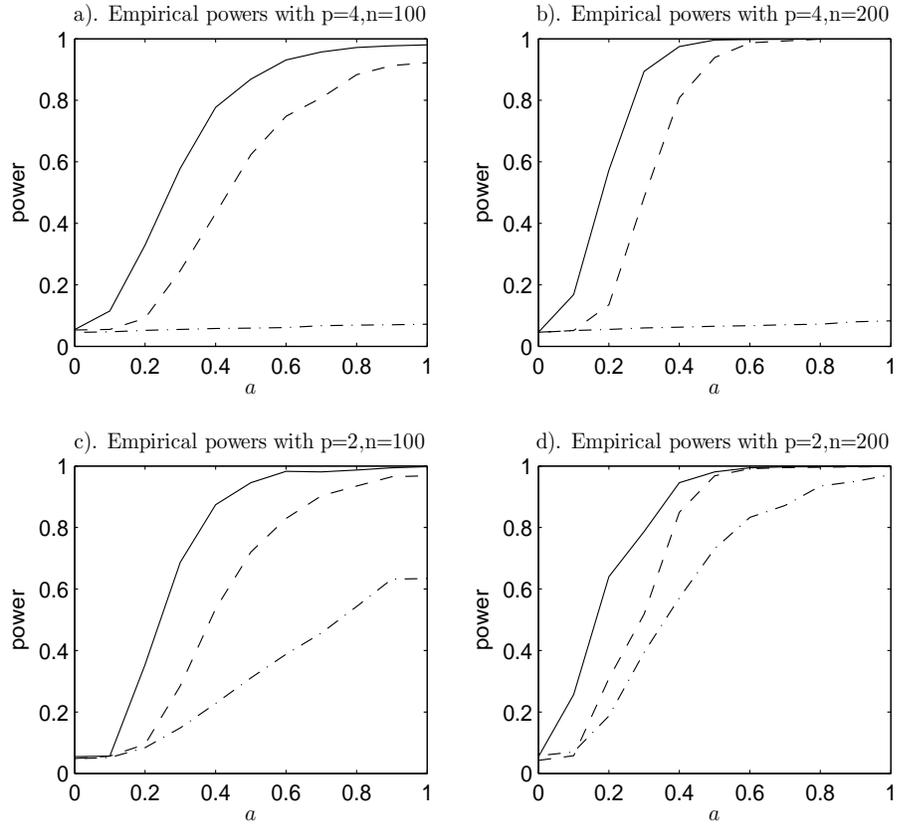}
 \caption{\small
 Empirical sizes and powers of $\tilde{S}_n^{OPG}$, $\tilde{S}_n^{DEE}$ and $T_n^{WQ}$ for $H_{20}$ v.s. $H_{23}$ at the significance level $\alpha=0.05$ with $X\sim N(0,I_p)$ and $\varepsilon\sim N(0,1)$.  In four plots, the solid line, the dash line and the dash-dotted line are for $\tilde{S}_n^{OPG}$, $\tilde{S}_n^{DEE}$ and $T_n^{WQ}$, respectively.
  }
  \label{plotH23}
\end{figure}

The simulation results with the alternatives $H_{21},H_{22}$ are reported in Table~\ref{tab_study2} for $a=0,0.2,\ldots,1$ at the significance level $\alpha=0.05$.
From this table, we can observe that for all of cases we conduct, the three tests can control empirical sizes very well. Also, it is reasonable that the simulated powers of all of tests become higher with increasing of the parameter $a$ and  the tests are more powerful with larger sample size. It can be seen clearly that the tests $\tilde{S}_n^{OPG}$ and $\tilde{S}_n^{DEE}$ work very well in  power performance. $T_n^{OPG}$ and $T_n^{DEE}$ are not significantly affected by the dimension of $X$. However, $T_n^{WQ}$ severely suffers from  the dimensionality problem. When the dimension of $X$ gets larger, $T_n^{WQ}$ completely fails to detect the alternatives. Figure~\ref{plotH23} reports the simulated power curves under $H_{23}$ for different values of $a$ and sample sizes $n=100,200$. The similar conclusion can be made.
\begin{center}
Table~\ref{tab_study2} and Figure~\ref{plotH23} about here
\end{center}

\vspace{0.2cm}
{\it Study~3}: We generate the data from the following models:
\begin{equation*}
    H_{31}: Y=\beta^\top X+2a(\beta^\top X)^3+\varepsilon,
\end{equation*}
where $\beta=(1,\ldots,1)^\top/\sqrt{p}$ and $p=8,12$. Here, the covariate $X=(X_1,\ldots,X_p)^\top$ come from the multivariate normal distribution $N(0,I_p)$ where $I_p$ is a $p\times p$ identity matrix and  $\varepsilon$ is from univariate standard normal distribution $N(0,1)$. In this study, we want to examine two issues: One is whether the test $T_n^{GWZ}$ can maintain empirical size when there exist some outlier values in responses and the other is whether  RDREAM can have an acceptable power performance when there are not outliers. The ratio of responses randomly replaced by observations from a nonlinear model $Y=6\exp(-|\beta^\top X|)+\varepsilon$ is denoted as $\rho$.

\begin{figure}[htbp]
\centering
\includegraphics[width=14cm,height=12cm]{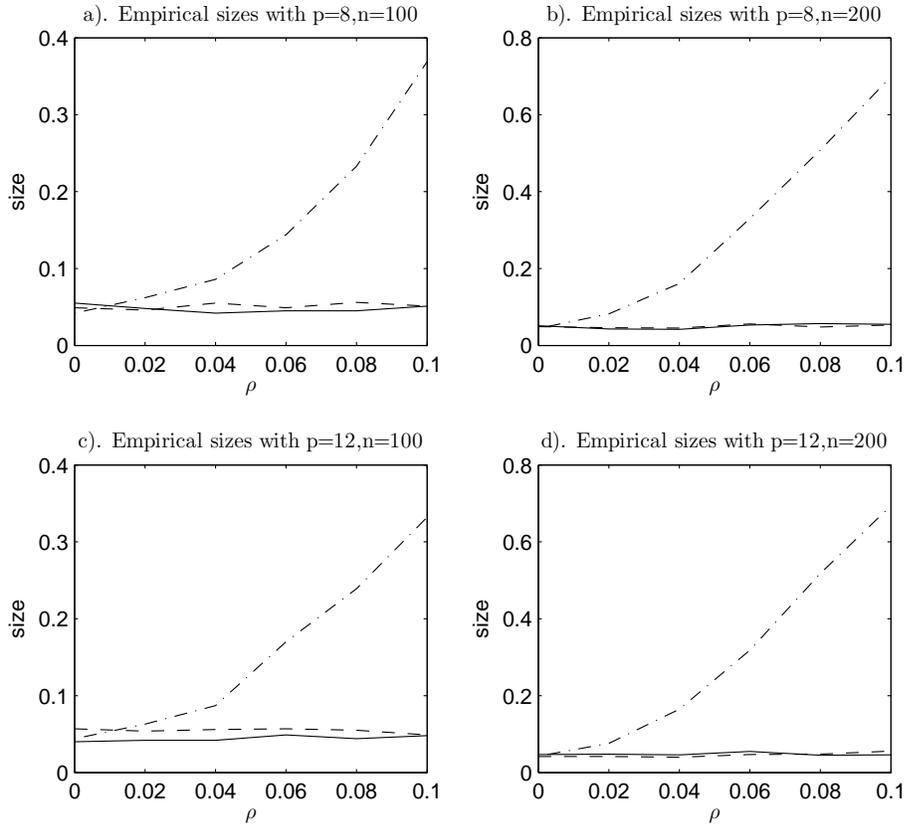}
 \caption{\small
 Empirical sizes of $\tilde{S}_n^{OPG}$, $\tilde{S}_n^{DEE}$ and $T_n^{GWZ}$ for $H_{30}$  at the significance level $\alpha=0.05$ with $X\sim N(0,I_p)$, $\varepsilon\sim N(0,1)$ and different values of $\rho$.  In four plots, the solid line, the dash line and the dash-dotted line are for $\tilde{S}_n^{OPG}$, $\tilde{S}_n^{DEE}$ and $T_n^{GWZ}$, respectively.
  }
  \label{plotH31}
\end{figure}

Figure~\ref{plotH31} presents the empirical size curves or ``significance trace'' of the three tests for different values of the ratio $\rho$ and sample sizes $n=100,200$. In this case, the ratio $\rho=0,0.02,\ldots,0.1$. From this figure, we can see that when $\rho=0$, all of the three tests can control empirical sizes very well which are all close to the pre-specified significance level 0.05. However, with the increasing of the ratio $\rho$, $\tilde{S}_n^{OPG}$ and $\tilde{S}_n^{DEE}$ outperform the test $T_n^{GWZ}$. The simulation results indicate that the new tests are not affected by the outlier values and they are more robust, whereas $T_n^{GWZ}$ fails to work when  outliers exist.
\begin{center}
Figure~\ref{plotH31} about here
\end{center}

\begin{figure}[htbp]
\centering
\includegraphics[width=14cm,height=12cm]{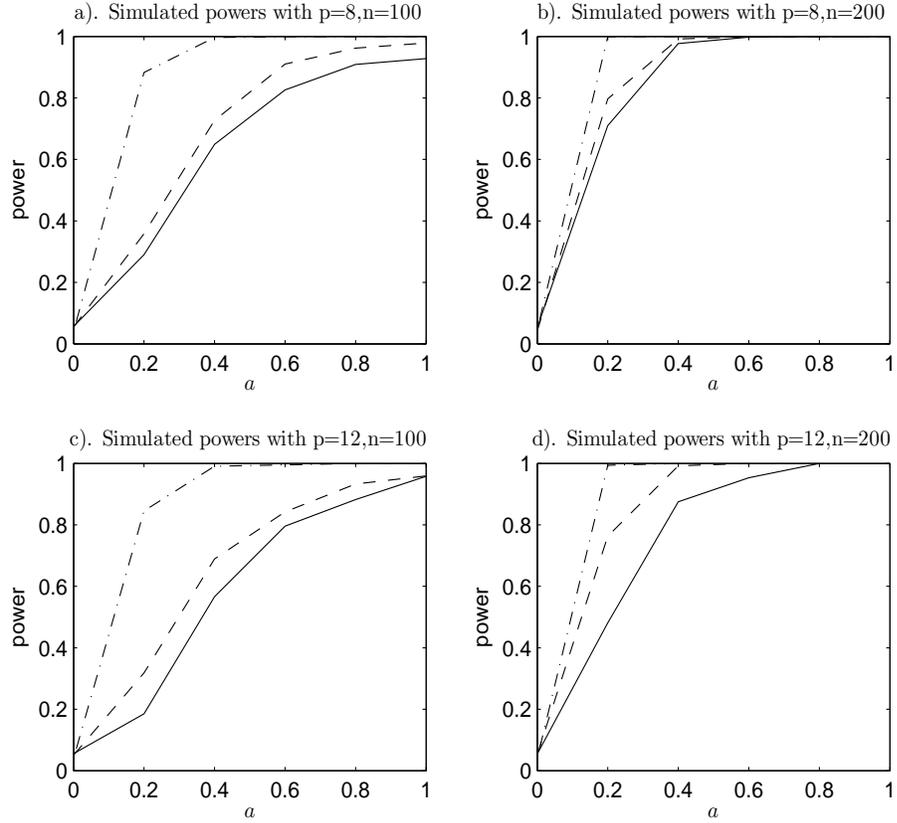}
 \caption{\small
 Simulated powers of $\tilde{S}_n^{OPG}$, $\tilde{S}_n^{DEE}$ and $T_n^{GWZ}$ for $H_{30}$ v.s. $H_{31}$ at the significance level $\alpha=0.05$ with $X\sim N(0,I_p)$ and $\varepsilon\sim N(0,1)$.  In four plots, the solid line, the dash line and the dash-dotted line are for $\tilde{S}_n^{OPG}$, $\tilde{S}_n^{DEE}$ and $T_n^{GWZ}$, respectively.
  }
  \label{plotH31_without}
\end{figure}

We display the simulated powers curves for different values of $a$ in Figure~\ref{plotH31_without}. Here, we consider $\rho=0$ and $n=100,200$. In other words, there are no outliers in the data. The parameter $a$ is set to be $0,0.2,\ldots,1$. Based on this figure, we can see that compared with $T_n^{GWZ}$, it is anticipated that $T_n^{GWZ}$ has  higher powers since their test employs more value-information whereas the robust tests only utilize the rank-information of responses. However, the  powers of the new tests are still  acceptable.
\begin{center}
Figure~\ref{plotH31_without} about here
\end{center}

\section{Real data analysis}
\renewcommand{\theequation}{6.\arabic{equation}}
\setcounter{equation}{0}
We now apply  RDREAM to a real data set collected from a HIV clinical trial. The HIV positive patients in this study were randomly divided into four groups to receive antiretroviral regimen: (i) ZDV; (ii) didanosine (ddi); (iii) ZDV+ddi and (iv) ZDV+zalcitabine. A more detailed description of this real data set can be found in Hammer et al. (1996). Many researchers have made use of this data set to illustrate their dimension reduction estimation methods and further to compare the treatment effects of monotherapy (say (i)) and combined therapy (say (ii)-(iv)), including Ding and Wang (2011), Guo et al (2014) and Hu et al (2010). Recently, Niu et al (2015a)
analyzed this data set to test whether the nonparametric component is a partial linear regression function.  The conclusion arrived was that a linear regression model would be proper for this data set.

In this dataset, there are 746 male patients who had not received antiretroviral therapy before the clinical trial and our study focuses on 473 patients who had observations in the variable CD4 cell counts at 96$\pm$5 weeks post therapy among them. Further, based on the way of therapy, we divide the dataset into two subsets: the first dataset has 105 male patients receiving monotherapy, say (i) and the second dataset contains 368 patients receiving combined therapies, say (ii)-(iv). For each dataset, the response variable $Y$ is CD4 cell counts at $96\pm5$ weeks post therapy and the four covariates are CD4 cell counts at baseline ($X_1$), CD4 cell counts at $20\pm5$ weeks ($X_2$), CD8 cell counts at baseline ($X_3$) and CD8 cell counts at $20\pm 5$ weeks ($X_4$). For ease of explanation, all the covariates are standardized separately and the responses are centered.

It is our interest to test whether the data $(Y,X)$ can be fitted with linear regression models where $X=(X_1,X_2,X_3,X_4)^\top$, that is,
\begin{eqnarray}
  &&H_{10}: E(Y|X)=\beta_1^\top X~\mbox{for some}~\beta_1\in R^4,\nonumber\\
  &&H_{11}: E(Y|X)=m(B_1^\top X)\neq \beta_1^\top X~\mbox{for any}~\beta_1\in R^4\label{realtest1}
\end{eqnarray}
for the first dataset and
\begin{eqnarray}
  &&H_{20}: E(Y|X)=\beta_2^\top X~\mbox{for some}~\beta_2\in R^4,\nonumber\\
  &&H_{21}: E(Y|X)=m(B_2^\top X)\neq \beta_2^\top X~\mbox{for any}~\beta_2\in R^4\label{realtest2}
\end{eqnarray}
for the second dataset, respectively. The same kernel function and bandwidth are adopted
as simulation section. With our proposed RDREAM, we can obtain that for the first dataset, the $p$-values for tests $\tilde{S}_n^{OPG}$ and $\tilde{S}_n^{DEE}$ are 0.696 and 0.678, respectively and the corresponding $p$-values for the second dataset are 0.645 and 0.361. All of these results indicate that the original two datasets can be fitted by linear regression models.

To investigate the influence of outlier values in the response space, we artificially replace the first $5\%$ responses of each dataset by an outlying value $800$. With these two new datasets, we apply our proposed methods to test (\ref{realtest1}) and (\ref{realtest2}), respectively. As to the first new dataset, the $p$-values for $\tilde{S}_n^{OPG}$ and $\tilde{S}_n^{DEE}$ are 0.546 and 0.780, respectively. The corresponding $p$-values are 0.543 and 0.151 for the second dataset. The similar conclusion with original data can be made, which indicates that our test methods are robust.

We next add $c_1=(3,4,5,6)$ copies of $(800,168,174,605,640)$ to the first original dataset and
$c_2=(12,15,18,21)$ copies of $(400,370,373,739,606)$ to the second original dataset, respectively, to carry out the same tests. Here, the maximum contamination rate is $5\%$. The $p$-values of these tests are listed in Table~\ref{tab_real}, which all suggest  linear regression modelling.
\begin{center}
Table~\ref{tab_real} about here
\end{center}

%

\section*{Appendix. Proofs of  theorems}
\renewcommand{\theequation}{A.\arabic{equation}}
\setcounter{equation}{0}

\noindent The following conditions are required for proving the theorems in Section
 \ref{sec3}.

\begin{itemize}
    \item [(C1)] The joint probability density function $f(z,y)$ of $(Z,Y)$ is bounded. Both errors $e_i$ and $\varepsilon_i$ have bounded probability density functions.

    \item [(C2)] The density function $f(B^\top X)$ of $B^\top X$ on support $\mathcal{Z}$ exists and has two bounded derivatives and satisfies
        $$0<\inf f(z)<\sup f(z)<1.$$

    \item [(C3)] The kernel function $\mathcal{K}(\cdot)$ is a bounded, derivative and symmetric probability density function  and all the moments of $\mathcal{K}(\cdot)$ exist. The bandwidth satisfies $nh^2\rightarrow \infty$, $nh^{\hat q}\rightarrow \infty$, $\int \mathcal{K}(u)du=1$.

    \item [(C4)] There exists an estimator $\hat\alpha$ such that under the null hypothesis, $\sqrt{n}(\hat\alpha-\alpha)=O_p(1)$, where $\alpha=(\beta,\theta)$ and under the local alternative sequences, $\sqrt{n}(\hat\alpha-\tilde\alpha)=O_p(1)$, where $\alpha$ and $\tilde\alpha$ are both interior points of $\Theta$, a compact and convex set.

    \item [(C5)] Denote $\alpha=(\beta,\theta)^\top$ and there exists a positive continuous function $G(x)$ such that $\forall \alpha_1,\alpha_2$, $|g(x,\alpha_1)-g(x,\alpha_2)|\leq G(x)|\alpha_1-\alpha_2|$.

    \item [(C6)] The matrix $E\{\nabla m(B^\top X)\nabla m(B^\top X)^\top\}$ is positive definite where $\nabla m(\cdot)=m'(\cdot)$ denotes the gradient of the function $m(\cdot)$.

    \item [(C7)] $\mathcal{M}_n(s)$ has the following expansion:
    \begin{equation*}
        \mathcal{M}_n(s)=\mathcal{M}(s)+E_n\{\psi(X,Y,s)\}+R_n(s),
    \end{equation*}
    where $E_n(\cdot)$ denotes the average over all sample points, $E\{\psi(X,Y,s)\}=0$ and $E\{\psi^2(X,Y,s)\}<\infty$.

    \item [(C8)] $\sup_s\parallel R_n(s)\parallel_F=o_p(n^{-1/2})$, where $\parallel\cdot\parallel_F$ denotes the Frobenius norm of a matrix.
\end{itemize}

\begin{remark}\label{remark1}
The conditions $(C1)$ and $(C5)$ are necessary for the robustness of Hampel influence function.
Conditions $(C2)-(C4)$ are needed for ensuring the asymptotic normality of our test statistic and the consistency of the parameter estimators, where
condition $(C3)$ is the common requisite for the kernel density estimation problem.
Condition $(C6)$ is assumed for OPG and $(C7)-(C8)$ are for DEE.
\end{remark}

\vspace{0.3cm}
The following lemmas are used to prove the theorems  in
Section~3. We first give the proof of Lemma~\ref{hatq_consist} in Section~2.

\vspace{0.3cm} \noindent \textit{Proof of Lemma~\ref{hatq_consist}.}
Under the null hypothesis $H_0$ and fixed alternative hypothesis $H_n$, the consistency for OPG-based estimate $\hat q\rightarrow q$ as $n\rightarrow\infty$ has been proved in Lemma~1 of Niu et al. (2015b). In the following, we only give the proof of DEE-based estimate and the same conditions in Theorem~4 of Zhu et al. (2010) are adopted.

Based on Theorem~2 in Zhu et al. (2010), under some conditions designed by them, it can be shown that $\mathcal{M}_{n,n}-\mathcal{M}=O_p(n^{-1/2})$. Further, the root-n consistency of the eigenvalues of $\mathcal{M}_{n,n}$ is retained, that is, $\hat\lambda_i-\lambda_i=O_p(n^{-1/2})$.
Note that when $l\leq q$, $\lambda_l>0$ and for $l>q$, we have $\lambda_l=0$. $c=1/\sqrt{nh}$ in our paper is recommended, thus, when $nh\rightarrow\infty, h\rightarrow 0$, we have $1/\sqrt n=o(c)$ and $c=o(1)$. For $1\leq l<q$,
\begin{eqnarray*}
\frac{\hat\lambda_{q+1}+c}{\hat\lambda_{q}+c}-\frac{\hat\lambda_{l+1}+c}{\hat\lambda_{l}+c}
&=&\frac{\lambda_{q+1}+c+O_p(\frac{1}{\sqrt n})}{\lambda_{q}+c+O_p(\frac{1}{\sqrt n})}-\frac{\lambda_{l+1}+c+O_p(\frac{1}{\sqrt n})}{\lambda_{l}+c+O_p(\frac{1}{\sqrt n})}
  \\
  &=&\frac{c+O_p(\frac{1}{\sqrt n})}{\lambda_{q}+c+O_p(\frac{1}{\sqrt n})}-\frac{\lambda_{l+1}+c+O_p(\frac{1}{\sqrt n})}{\lambda_{l}+c+O_p(\frac{1}{\sqrt n})}
  \Rightarrow -\frac{\lambda_{l+1}}{\lambda_{l}}<0.
\end{eqnarray*}
When $l>q$,
\begin{eqnarray*}
  \frac{\hat\lambda_{q+1}+c}{\hat\lambda_{q}+c}-\frac{\hat\lambda_{l+1}+c}{\hat\lambda_{l}+c}
  &=&\frac{\lambda_{q+1}+c+O_p(\frac{1}{\sqrt n})}{\lambda_{q}+c+O_p(\frac{1}{\sqrt n})}-\frac{\lambda_{l+1}+c+O_p(\frac{1}{\sqrt n})}{\lambda_{l}+c+O_p(\frac{1}{\sqrt n})}
  \\
  &=&\frac{c+O_p(\frac{1}{\sqrt n})}{\lambda_{q}+c+O_p(\frac{1}{\sqrt n})}-\frac{c+O_p(\frac{1}{\sqrt n})}{c+O_p(\frac{1}{\sqrt n})}\Rightarrow -1<0.
\end{eqnarray*}
Therefore, we can conclude that under the null hypothesis $H_0$ and under the fixed alternative hypothesis (\ref{alter_model}), $\hat q\rightarrow q$ as $n\rightarrow\infty$, which completes the proof. $\Box$

\vspace{0.3cm}
The proof of Lemma~\ref{consis_qlocal} in Section~3 is given as follows.

\vspace{0.3cm} \noindent \textit{Proof of Lemma~\ref{consis_qlocal}.}
Under the local alternative hypothesis (\ref{loca_alter}) with $C_n=n^{-1/2}h^{-1/4}$, the proof of OPG-based $\hat q\rightarrow 1$ as $n\rightarrow\infty$ has been given in Lemma~1 of Niu et al (2015b). We only state the proof for DEE-based estimate.

From the proof of Theorem~2 in Guo et al (2015), it is shown that under the local alternative hypothesis, $\mathcal{M}_{n,n}-\mathcal{M}=O_p(C_n)$. Further, we can get
$\hat\lambda_i-\lambda_i=O_p(C_n)$.

Thus, Note that
$\lambda_1>0$ and for any $l > 1$, we have $\lambda_l = 0$.
Consequently, under the condition that $C_n=o(c)$ and $c=o(1)$,
\begin{eqnarray*}
\frac{\hat\lambda_2+c}{\hat\lambda_1+c}-\frac{\hat\lambda_{l+1}+c}{\hat\lambda_l+c}&=&
\frac{\lambda_2+c+O_p(C_n)}{\lambda_1+c+O_p(C_n)}-\frac{\lambda_{l+1}+c+O_p(C_n)}{\lambda_l+c+O_p(C_n)}\\
&=&\frac{c+O_p(C_n)}{\lambda_1+c+O_p(C_n)}-\frac{c+O_p(C_n)}{c+O_p(C_n)}\Rightarrow
-1<0.
\end{eqnarray*}
Thus under the local alternative (\ref{loca_alter}), Lemma 1 holds. The proof of Lemma~\ref{consis_qlocal} is finished. $\Box$

\vspace{0.3cm}
\begin{lemma}\label{consis_Vn}
Given the conditions (C1)-(C8) in the Appendix,
we have
\begin{equation*}
    nh^{1/2}(V_n-V_n^\star)\stackrel{p}{\rightarrow}0,
\end{equation*}
where $\stackrel{p}{\rightarrow}$ represents convergence in probability and
\begin{equation}\label{Vnstar}
    V_n^\star=\frac{1}{n(n-1)} \sum_{i=1}^n\sum_{j\neq i}^n \mathcal{K}_h\{\hat B(\hat q)^\top (x_i-x_j)\}[H(e_i)-\frac{1}{2}][H(e_j)-\frac{1}{2}],
\end{equation}
here, $\mathcal{K}_h(\cdot)=\mathcal{K}(\cdot/h)/h^{\hat q}$.
\end{lemma}

\vspace{0.3cm}
\noindent \textit{Proof of Lemma~\ref{consis_Vn}.}
We first decompose $V_n-V_n^\star$ as
\begin{eqnarray}
  V_n-V_n^\star&=&\frac{1}{n^3(n-1)} \sum_{i=1}^n\sum_{j\neq i}^n\sum_{l=1}^n\sum_{k=1}^n\mathcal{K}_h\{\hat B(\hat q)^\top (x_i-x_j)\}\nonumber\\
  &&\times[I(\hat e_l\leq \hat e_i)-I(e_l\leq e_i)]
[I(\hat e_k\leq \hat e_j)-I(e_k\leq e_j)]\nonumber\\
  &&+\frac{2}{n^3(n-1)} \sum_{i=1}^n\sum_{j\neq i}^n\sum_{l=1}^n\sum_{k=1}^n\mathcal{K}_h\{\hat B(\hat q)^\top (x_i-x_j)\}\nonumber\\
  &&\times[I(\hat e_l\leq \hat e_i)-I(e_l\leq e_i)]
[I(e_k\leq e_j)-\frac{n+1}{2n}]\nonumber\\
&&+\frac{1}{n^3(n-1)} \sum_{i=1}^n\sum_{j\neq i}^n\sum_{l=1}^n\sum_{k=1}^n\mathcal{K}_h\{\hat B(\hat q)^\top (x_i-x_j)\}\nonumber\\
  &&\times\Big\{[I(e_l\leq e_i)-\frac{n+1}{2n}][I(e_k\leq e_j)-\frac{n+1}{2n}]\nonumber\\
  &&-[H(e_i)-\frac{1}{2}][H(e_j)-\frac{1}{2}]\Big\}\nonumber\\
  &=:&A_1+A_2+A_3.\label{Vn_Vnstar}
\end{eqnarray}

Since $e_i=y_i-g(\beta^\top x_i,\theta)$, we further have $\hat e_i-e_i=g(\beta^\top x_i,\theta)-g(\hat{\beta}^\top x_i,\hat{\theta})$. Let $\alpha=(\beta,\theta)^\top$ and $L(\hat\alpha)$ as
\begin{equation*}
    L(\hat\alpha)=\max_{1\leq i\leq n}\sum_{l=1,l\neq i}^n |I(\hat e_l\leq \hat e_i)-I(e_l\leq e_i)|.
\end{equation*}
Denote $\Omega=\{\alpha^\star: \sqrt n|\alpha^\star-\alpha_0|\leq \delta\}$ for $\delta=O(1)$ and
$t(x_i,x_l,\alpha,\alpha^\star)=[g(\beta^\top x_i,\theta)-g({\beta^\star}^\top x_i,\theta^\star)]-[g({\beta}^\top x_l,\theta)-g({\beta^\star}^\top x_l,{\theta^\star})]$, then
\begin{eqnarray}
  \sup_{\alpha^\star\in\Omega}|L(\alpha^\star)|
  &=&\sup_{\alpha^\star\in\Omega}\sum_{l=1,l\neq i}^n|I(e_l-e_i\leq t(x_i,x_l,\alpha,\alpha^\star))-I(e_l\leq e_i)|\nonumber\\
  &\leq&\sup_{\alpha^\star\in\Omega}\sum_{l=1,l\neq i}^n I(|e_l-e_i|\leq |t(x_i,x_l,\alpha,\alpha^\star)|)\nonumber\\
  &\leq&\sum_{l=1,l\neq i}^n I(|e_l-e_i|\leq Cn^{-1/2}),\label{supL}
\end{eqnarray}
where $C$ is a generic positive constant. From the above derivation, we can obtain that, conditional on $e_i$, $I(|e_l-e_i|\leq Cn^{-1/2}), l\neq i$ are iid Bernoulli random variables with $O(n^{-1/2})$ order success probability. Using Bernstein's inequality leads to $P\{\sum_{l=1,l\neq i}^n I(|e_l-e_i|\leq Cn^{-1/2})\geq Cn^{1/2}|e_i\}\leq \exp(-Cn^{1/2})$. Unconditionally, we still have that
\begin{equation*}
    P\Big\{\sum_{l=1,l\neq i}^n I(|e_l-e_i|\leq Cn^{-1/2})\geq Cn^{1/2}\Big\}\leq \exp(-Cn^{1/2}).
\end{equation*}
Further,
\begin{equation*}
    P\Big\{\max_{1\leq i\leq n}\sum_{l=1,l\neq i}^n I(|e_l-e_i|\leq Cn^{-1/2})\geq Cn^{1/2}\Big\}\leq n\exp(-Cn^{1/2}).
\end{equation*}
When $n\rightarrow\infty$, for a positive constant $C$, $n\exp(-Cn^{1/2})\rightarrow 0$. Therefore, $P\{\max_{1\leq i\leq n}\sum_{l=1,l\neq i}^n I(|e_l-e_i|\leq Cn^{-1/2})< Cn^{1/2}\}=1$. Then,
\begin{equation}\label{probab_bound}
    \max_{1\leq i\leq n}\sum_{l=1,l\neq i}^n I(|e_l-e_i|\leq Cn^{-1/2})< Cn^{1/2}=O_p(n^{1/2}).
\end{equation}
Combining (\ref{supL}) and (\ref{probab_bound}), we can gain the following useful probability bound
\begin{equation*}
    L(\hat\alpha)=\max_{1\leq i\leq n}\sum_{l=1,l\neq i}^n |I(\hat e_l\leq \hat e_i)-I(e_l\leq e_i)|=O_p(n^{1/2}).
\end{equation*}

An application of the formula (\ref{probab_bound}) and for the term $A_1$ in (\ref{Vn_Vnstar}), we have
\begin{eqnarray}
  nh^{1/2}|A_1|&\leq&\frac{h^{1/2}}{n^2(n-1)h^{\hat q}}\max_{1\leq i\leq n}\sum_{l=1,l\neq i}^n |I(\hat e_l\leq \hat e_i)-I(e_l\leq e_i)|\nonumber\\
  &&\times \max_{1\leq j\leq n}\sum_{k=1,k\neq j}^n |I(\hat e_k\leq \hat e_j)-I(e_k\leq e_j)|\nonumber\\
  &&\times \sum_{i=1}^n\sum_{j\neq i}^n \mathcal{K}\{\hat B(\hat q)^\top (x_i-x_j)/h\}\nonumber\\
  &=&\frac{Ch^{1/2}}{n(n-1)h^{\hat q}}\sum_{i=1}^n\sum_{j\neq i}^n \mathcal{K}\{\hat B(\hat q)^\top (x_i-x_j)/h\}\nonumber\\
  &=&\frac{Ch^{1/2}}{n(n-1)h^{\hat q}}\sum_{i=1}^n\sum_{j\neq i}^n \mathcal{K}\{B^\top (x_i-x_j)/h\}\nonumber\\
  &&+\frac{Ch^{1/2}}{n(n-1)h^{\hat q}}\sum_{i=1}^n\sum_{j\neq i}^n\Big[\mathcal{K}\{\hat B(\hat q)^\top (x_i-x_j)/h\}-\mathcal{K}\{B^\top (x_i-x_j)/h\}\Big]\nonumber\\
  &=:&C_1(A_{11}+A_{12}),\label{A11A12}
\end{eqnarray}
where $C_1$ is a positive constant. Let $Z=B^\top X$. As to the term $A_{11}$, the term
\begin{equation*}
    \frac{1}{n(n-1)h^{\hat q}}\sum_{i=1}^n\sum_{j\neq i}^n \mathcal{K}\{B^\top (x_i-x_j)/h\}
\end{equation*}
is an U-statistic with the kernel as
$H_n(z_1,z_2)=h^{-\hat q}\mathcal{K}\{(z_1-z_2)/h\}$. In order to apply the theory for non-degenerate U-statistic (Serfling 1980), $E[H_n(z_1,z_2)^2]=o(n)$ is needed. It can be verified that
\begin{eqnarray}\label{EHsquare}
  E[H_n(z_1,z_2)^2]&=&E\{E[H_n(z_1,z_2)^2|z_1,z_2]\}\nonumber\\
  &=&\int \frac{1}{h^{2\hat q}}\mathcal{K}^2(\frac{z_1-z_2}{h})p(z_1)p(z_2)dz_1dz_2\nonumber\\
  &=&\int \frac{1}{h^{2\hat q}}\mathcal{K}^2(u)p(z_1)p(z_1-hu)(-h^{\hat q})dz_1du\nonumber\\
  &=&-\frac{1}{h^{\hat q}}\int\mathcal{K}^2(u)p^2(z_1)dz_1du+o(1)\nonumber\\
  &=&O(\frac{1}{h^{\hat q}}),
\end{eqnarray}
where $p(\cdot)$ is denoted as the probability density function.
With the condition $nh^{\hat q}\rightarrow\infty$, we have $E[H_n(z_1,z_2)^2]=O(1/h^{\hat q})=o(n)$.
The condition of lemma~3.1 of Zheng (1996) is satisfied and we have
$A_{11}=h^{1/2}E[H_n(z_1,z_2)]+o_p(1)$,
where $E[H_n(z_1,z_2)]=O(1)$. Therefore we can obtain
that $A_{11}=O_p(h^{1/2})=o_p(1)$. Denote
\begin{equation*}
    A_{12}^\star=\frac{1}{n(n-1)}\sum_{i=1}^n\sum_{j\neq i}^n h^{1/2-\hat q}
    \mathcal{K}'\{\tilde B^\top (x_i-x_j)/h\}(x_i-x_j)^\top\times\frac{\hat B(\hat q)-B}{h},
\end{equation*}
where $\tilde{B}$ lies between $B$ and $\hat B$. Then for the term $A_{12}$ in (\ref{A11A12}), we have
\begin{equation*}
    A_{12}=A_{12}^\star+o_p(A_{12}^\star).
\end{equation*}
Similar to $A_{11}$, the following term
\begin{equation*}
    \frac{1}{n(n-1)}\sum_{i=1}^n\sum_{j\neq i}^n h^{-\hat q}
    \mathcal{K}'\{\tilde B^\top (x_i-x_j)/h\}(x_i-x_j)^\top
\end{equation*}
can be regarded as an U-statistic. It can be similarly shown that the term is the order of $O_p(h)$. As $\|\hat B(\hat q)-B\|_2=O_p(1/\sqrt n)$ and under the condition $n\rightarrow\infty, h\rightarrow 0$, we can obtain that $A_{12}=o_p(1)$. Towards to (\ref{A11A12}), we have
\begin{equation*}
    nh^{1/2}|A_1|\leq o_p(1).
\end{equation*}
Similarly, we can derive that $nh^{1/2}|A_i|=o_p(1)$ for $i=2,3$. Combining with the formula (\ref{Vn_Vnstar}), it can be concluded that
\begin{equation*}
    nh^{1/2}(V_n-V_n^\star)\stackrel{p}{\rightarrow}0,
\end{equation*}
which completes the proof of Lemma~\ref{consis_Vn}. $\Box$

\vspace{0.3cm}
In the following, we give the proof for Theorem~\ref{theo_null}.

\noindent \textit{Proof of Theorem~\ref{theo_null}.}
From Lemma~{\ref{consis_Vn}}, we know that the limiting distributions for $nh^{1/2}V_n$ and $nh^{1/2}V_n^\star$ are the same. Thus, we just need to derive the asymptotic property of $nh^{1/2}V_n^\star$. The term $V_n^\star$ in (\ref{Vnstar}) can be decomposed as
\begin{eqnarray*}
  V_n^\star&=&\frac{1}{n(n-1)} \sum_{i=1}^n\sum_{j\neq i}^n \mathcal{K}_h\{B^\top (x_i-x_j)\}[H(e_i)-\frac{1}{2}][H(e_j)-\frac{1}{2}]\\
  &&+\frac{1}{n(n-1)} \sum_{i=1}^n\sum_{j\neq i}^n [H(e_i)-\frac{1}{2}][H(e_j)-\frac{1}{2}]\Big[\mathcal{K}_h\{\hat B(\hat q)^\top (x_i-x_j)\}
  -\mathcal{K}_h\{B^\top (x_i-x_j)\}\Big]\\
  &=:&V_{n1}^\star+V_{n2}^\star,
\end{eqnarray*}
where $\mathcal{K}_h(\cdot)=\mathcal{K}(\cdot/h)/h^{\hat q}$.

For the term $V_{n1}^\star$, it is a U-statistic, since we always assume that the dimension of $B^\top X$ is fixed in our paper. Under the null hypothesis, $H(e_i),\,\,i=1,\ldots,n$ follows a uniform distribution on $(0,1)$, $q=1$ and $\hat q\rightarrow 1$. An application of Theorem~1 in Zheng (1996), it is not difficult to derive the asymptotic normality: $nh^{1/2}V_{n1}^\star\Rightarrow N(0,Var)$, where
\begin{equation*}
    Var=\frac{1}{72}\int \mathcal{K}^2(u)du\int p^2(z)dz
\end{equation*}
with $Z=B^\top X$.

Denote
\begin{equation*}
    \tilde V_{n2}^\star=\frac{1}{n(n-1)} \sum_{i=1}^n\sum_{j\neq i}^n \frac{1}{h^{\hat q}}\mathcal{K}'\Big\{\frac{\tilde{B}^\top (x_i-x_j)}{h}\Big\}(x_i-x_j)^\top[H(e_i)-\frac{1}{2}][H(e_j)-\frac{1}{2}]\cdot\frac{\hat B(\hat q)-B}{h},
\end{equation*}
where $\tilde B$ lies between $\hat B$ and $B$. An application of Taylor expansion yields
\begin{equation*}
    V_{n2}^\star=\tilde V_{n2}^\star+o_p(\tilde V_{n2}^\star).
\end{equation*}
Because the kernel $\mathcal{K}(\cdot)$ is spherical symmetric, the following term can be considered as an U-statistic:
\begin{equation*}
    \frac{1}{n(n-1)} \sum_{i=1}^n\sum_{j\neq i}^n \frac{1}{h^{\hat q}}\mathcal{K}'\{\tilde{B}^\top (x_i-x_j)/h\}(x_i-x_j)^\top[H(e_i)-\frac{1}{2}][H(e_j)-\frac{1}{2}].
\end{equation*}
Further note that
\begin{eqnarray*}
  &&E\Big\{\frac{1}{h^{\hat q}}\mathcal{K}'\{\tilde{B}^\top (x_i-x_j)/h\}(x_i-x_j)^\top[H(e_i)-\frac{1}{2}][H(e_j)-\frac{1}{2}]|x_i,y_i\Big\}\\
  &=&E\left[E\Big\{\frac{1}{h^{\hat q}}\mathcal{K}'\{\tilde{B}^\top (x_i-x_j)/h\}(x_i-x_j)^\top[H(e_i)-\frac{1}{2}][H(e_j)-\frac{1}{2}]|x_i,y_i,x_j\Big\}
  |x_i,y_i\right]\\
  &=&E\left\{\frac{1}{h^{\hat q}}\mathcal{K}'\{\tilde{B}^\top (x_i-x_j)/h\}(x_i-x_j)^\top[H(e_i)-\frac{1}{2}]\cdot E[H(e_j)-\frac{1}{2}|x_j]|x_i,y_i\right\}=0.
\end{eqnarray*}
Thus the above U-statistic is degenerate. Similar as the derivation of $V_{n1}^\star$, together with $\|\hat B(\hat q)-B\|_2=O_p(1/\sqrt n)$ and $1/nh^2\rightarrow 0$, we have $nh^{1/2}V_{n2}^\star=o_p(1)$. Therefore, under the null hypothesis $H_0$, we can conclude that
$nh^{1/2}V_n^\star\Rightarrow N(0,Var)$. Based on Lemma~\ref{consis_Vn}, we have $nh^{1/2}V_n\Rightarrow N(0,Var)$.

An estimate of $Var$ can be defined as
\begin{equation*}
    \widehat{Var}=\frac{1}{72n(n-1)}\sum_{i=1}^n\sum_{j\neq i}^n \frac{1}{h^{\hat q}}\mathcal{K}^2\Big\{\frac{\hat B(\hat q)^\top (x_i-x_j)}{h}\Big\}.
\end{equation*}
Since the proof is rather straightforward, we then only give a brief description. Using a similar argument as that for Lemma~\ref{consis_Vn}, we can get
\begin{equation*}
    \widehat{Var}=\frac{1}{72n(n-1)}\sum_{i=1}^n\sum_{j\neq i}^n \frac{1}{h^{\hat q}}\mathcal{K}^2\Big\{\frac{B^\top (x_i-x_j)}{h}\Big\}+o_p(1).
\end{equation*}
The consistency can be derived through U-statistic theory.
The proof for Theorem~\ref{theo_null} is finished. $\Box$

\vspace{0.3cm}
The proof for Theorem~\ref{theo_alter} is given as follows.

\noindent \textit{Proof of Theorem~\ref{theo_alter}.}
Under the global alternative $H_{n}$ in (\ref{alter_model}), we have $e_i=m(B^\top x_i)+\varepsilon_i-g(\tilde{\beta}^\top x_i,\tilde{\theta})$.
Together with Lemma~\ref{consis_Vn}, it can be obtained that $V_n=V_n^\star+o_p(1)$,
where $V_n^\star$  can be rewritten as
\begin{eqnarray*}
  V_n^\star&=&\frac{1}{n(n-1)} \sum_{i=1}^n\sum_{j\neq i}^n \frac{1}{h^{\hat q}}\mathcal{K}\Big\{\frac{\hat B(\hat q)^\top (x_i-x_j)}{h}\Big\}
  \Big[H(m(B^\top x_i)+\varepsilon_i-g(\tilde{\beta}^\top x_i,\tilde{\theta}))-\frac{1}{2}\Big]\\
  &&\Big[H(m(B^\top x_j)+\varepsilon_j-g(\tilde{\beta}^\top x_j,\tilde{\theta}))-\frac{1}{2}\Big]\\
  &=&\frac{1}{n(n-1)} \sum_{i=1}^n\sum_{j\neq i}^n \frac{1}{h^{\hat q}}\mathcal{K}\Big\{\frac{ B^\top (x_i-x_j)}{h}\Big\}
  \Big[H(m(B^\top x_i)+\varepsilon_i-g(\tilde{\beta}^\top x_i,\tilde{\theta}))-\frac{1}{2}\Big]\\
  &&\Big[H(m(B^\top x_j)+\varepsilon_j-g(\tilde{\beta}^\top x_j,\tilde{\theta}))-\frac{1}{2}\Big]+\frac{1}{n(n-1)} \sum_{i=1}^n\sum_{j\neq i}^n\Big[H(m(B^\top x_i)+\varepsilon_i\\
  &&-g(\tilde{\beta}^\top x_i,\tilde{\theta}))-\frac{1}{2}\Big]
  \Big[H(m(B^\top x_j)+\varepsilon_j-g(\tilde{\beta}^\top x_j,\tilde{\theta}))-\frac{1}{2}\Big]\Big[\mathcal{K}_h\{\hat B(\hat q)^\top (x_i-x_j)\}\\
  &&-\mathcal{K}_h\{B^\top (x_i-x_j)\}\Big]\\
  &=&V_{n3}^\star+V_{n4}^\star,
\end{eqnarray*}

For the term $V_{n3}^\star$, it is a standard U-statistic with
\begin{equation*}
    H_n(x_i,x_j)=\frac{1}{h^{\hat q}}\mathcal{K}\Big\{\frac{ B^\top (x_i-x_j)}{h}\Big\}l(x_i)l(x_j),
\end{equation*}
where $l(x_\cdot)=[H\{m(B^\top x_\cdot)+\varepsilon_\cdot-g(\tilde{\beta}^\top x_\cdot,\tilde{\theta})\}-1/2]$. Similar to the proof of (\ref{EHsquare}), when $nh^{\hat q}\rightarrow\infty$, we can derive that $E[H^2(x_i,x_j)]=o(n)$ and the condition of Lemma~3.1 in Zheng (1996) can be shown to be satisfied. We further cacluate
\begin{eqnarray*}
  E[H_n(x_i,x_j)]&=&E\{E[H_n(x_i,x_j)|x_i,x_j]\}\\
  &=&\frac{1}{h^{\hat q}}\int \mathcal{K}\Big\{\frac{z_i-z_j}{h}\Big\}\tilde{l}(z_i)\tilde{l}(z_j)p(z_i)
  p(z_j)dz_i dz_j\\
  &=&\frac{1}{h^{\hat q}}\int \mathcal{K}(u)\tilde{l}(z_j+hu)\tilde{l}(z_j)p(z_j+hu)p(z_j)\times
  h^{\hat q}du dz_j\\
  &=&\int \tilde{l}^2(z_j)p^2(z_j)dz_j+o(1)\\
  &=&E[l^2(X)^2p(X)]+o(1).
\end{eqnarray*}
where $Z=B^\top X$. Therefore, $V_{n3}^\star=E[l^2(X)^2p(X)]+o_p(1)=:C_2$, here,
$C_2$ is a positive constant.

As to the term $V_{n4}^\star$, similarly as the term $V_{n2}^\star$, we have
\begin{equation*}
    V_{n4}^\star=\tilde V_{n4}^\star+o_p(\tilde V_{n4}^\star),
\end{equation*}
where,
\begin{equation*}
    \tilde V_{n4}^\star=\frac{1}{n(n-1)} \sum_{i=1}^n\sum_{j\neq i}^n \frac{1}{h^{\hat q}}\mathcal{K}'\Big\{\frac{\tilde{B}^\top (x_i-x_j)}{h}\Big\}(x_i-x_j)^\top l(x_i)l(x_j)\cdot\frac{\hat B(\hat q)-B}{h},
\end{equation*}
here, $\tilde B$ lies between $B$ and $\hat B$.
Similarly as the derivation of $V_{n3}^\star$, together with $\|\hat B(\hat q)-B\|_2=O_p(1/\sqrt n)$, when $nh^{\hat q}\rightarrow\infty$, we have $V_{n4}^\star=O_p(h)\cdot O_p(1/\sqrt n)\cdot (1/h)=o_p(1)$.

Based on the above analysis, we can derive that $V_n=C_2+o_p(1)$ and $nh^{1/2}V_n\Rightarrow\infty$ in probability, which completes the proof of the global alternative situation.

\vspace{0.3cm}
We now consider the situation of local alternative $H_{1n}$ in (\ref{loca_alter}). Based on Lemma~\ref{consis_Vn}, we have $V_n=V_n^\star+o_p(1)$. In this situation, $e_i=C_n m(B^\top x_i)+\varepsilon_i$. Therefore, $V_n^\star$ can be decomposed as
\begin{eqnarray}
  V_n^\star&=&\frac{1}{n(n-1)} \sum_{i=1}^n\sum_{j\neq i}^n \frac{1}{h^{\hat q}}\mathcal{K}\Big\{\frac{\hat B(\hat q)^\top (x_i-x_j)}{h}\Big\}
  \Big[H(C_n m(B^\top x_i)+\varepsilon_i)-\frac{1}{2}\Big]\nonumber\\
  &&\Big[H(C_n m(B^\top x_j)+\varepsilon_j)-\frac{1}{2}\Big]\nonumber\\
  &=&\frac{1}{n(n-1)} \sum_{i=1}^n\sum_{j\neq i}^n \frac{1}{h^{\hat q}}\mathcal{K}\Big\{\frac{ B^\top (x_i-x_j)}{h}\Big\}
  \Big[H(C_n m(B^\top x_i)+\varepsilon_i)-\frac{1}{2}\Big]\nonumber\\
  &&\Big[H(C_n m(B^\top x_j)+\varepsilon_j)-\frac{1}{2}\Big]+\frac{1}{n(n-1)} \sum_{i=1}^n\sum_{j\neq i}^n\Big[H(C_n m(B^\top x_i)+\varepsilon_i)\nonumber\\
  &&-\frac{1}{2}\Big]
  \Big[H(C_n m(B^\top x_j)+\varepsilon_j)-\frac{1}{2}\Big]\Big[\mathcal{K}_h\{\hat B(\hat q)^\top (x_i-x_j)\}-\mathcal{K}_h\{B^\top (x_i-x_j)\}\Big]\nonumber\\
  &=&V_{n5}^\star+V_{n6}^\star,\label{Vnstar_loca}
\end{eqnarray}

For the term $V_{n5}^\star$, taking a Taylor expansion of $H(C_n m(B^\top x_i)+\varepsilon_i)$ around $C_n=0$, we have
\begin{equation*}
    H(C_n m(B^\top x_i)+\varepsilon_i)=H(\varepsilon_i)+C_n h(\varepsilon_i)m(B^\top x_i)+o_p(C_n^2).
\end{equation*}
Then, the term $V_{n5}^\star$ can be decomposed as
\begin{eqnarray*}
V_{n5}^\star&=&\frac{1}{n(n-1)} \sum_{i=1}^n\sum_{j\neq i}^n \frac{1}{h^{\hat q}}\mathcal{K}\Big\{\frac{ B^\top (x_i-x_j)}{h}\Big\}
  \Big[H(\varepsilon_i)-\frac{1}{2}\Big]
  \Big[H(\varepsilon_j)-\frac{1}{2}\Big]\\
  &&+2C_n\left\{\frac{1}{n(n-1)} \sum_{i=1}^n\sum_{j\neq i}^n \frac{1}{h^{\hat q}}\mathcal{K}\Big\{\frac{ B^\top (x_i-x_j)}{h}\Big\}\Big[H(\varepsilon_i)-\frac{1}{2}\Big]
  h(\varepsilon_j)m(B^\top x_j)
  \right\}\\
  &&+C_n^2\left\{\frac{1}{n(n-1)} \sum_{i=1}^n\sum_{j\neq i}^n \frac{1}{h^{\hat q}}\mathcal{K}\Big\{\frac{ B^\top (x_i-x_j)}{h}\Big\}h(\varepsilon_i)m(B^\top x_i)h(\varepsilon_j)m(B^\top x_j)\right\}+o_p(C_n^2)\\
  &=&D_1+2C_n D_2+C_n^2D_3+o_p(C_n^2).
\end{eqnarray*}
Under the local alternative hypothesis, $\hat q\rightarrow 1$ can be obtained. For the term $D_1$, similarly to the proof of the term $V_{n1}^\star$ in Theorem~\ref{theo_null}, we can show that $nh^{1/2}D_1\Rightarrow N(0,Var)$, where
\begin{equation*}
    Var=\frac{1}{72}\int \mathcal{K}^2(u)du\int p^2(z)dz
\end{equation*}
with $Z=B^\top X$. As to the term $D_2$, similarly as the proof of Lemma~3.3b in Zheng (1996), it can be obtained that $D_2=O_p(1/\sqrt n)$. When $C_n=n^{-1/2}h^{-1/4}$, we have $nh^{1/2}C_nD_2=O_p(h^{1/2})$. Turn to the term $D_3$, similarly to the proof of $V_{n3}^\star$ in our Theorem~\ref{theo_alter}, we have
$D_3=E[h^2(\varepsilon)m^2(B^\top X)p(X)]+o_p(1)$. Further, $nh^{1/2}C_n^2D_3=E[h^2(\varepsilon)m^2(B^\top X)p(X)]+o_p(1)$. Therefore,
\begin{equation*}
    nh^{1/2}V_{n5}^\star\Rightarrow N(\mu,Var),
\end{equation*}
where $\mu=E[h^2(\varepsilon)m^2(B^\top X)p(X)].$

As to the term $V_{n6}^\star$, just similarly as the proof of the term $V_{n2}^\star$ in our Theorem~\ref{theo_null}, it can be gotten that $nh^{1/2}V_{n6}^\star=o_p(1)$.

Combining Lemma~\ref{consis_Vn} and the formula (\ref{Vnstar_loca}), under the local alternative, we have $nh^{1/2}V_n\Rightarrow N(\mu,Var)$.

The proof of Theorem~\ref{theo_alter} is finished. $\Box$


\vspace{0.3cm}
The verification for the formula (\ref{sec_deri}) is as follows.

\noindent \textit{Verification of (\ref{sec_deri}).}
Let $f_t$ and $u$ be the probability density functions of $F_t$ and $U$. Recall $F_t=F+tU$, then $dF_t=dF+t dU$ and $f_t=f+tu$,
we further have
\begin{eqnarray*}
  L(t)&=&\int\int\Big[H(y-g(\beta^\top x(F_t),\theta(F_t)))-\frac{1}{2}\Big]\times\Big(
  \int\Big[H(y_1-g(\beta^\top x(F_t),\theta(F_t)))-\frac{1}{2}\Big]\nonumber\\
  &&\times f_t(z,y_1)dy_1\Big)d F_t(z,y),\\
  &=&\int\int\Big[H(y-g(\beta^\top x(F_t),\theta(F_t)))-\frac{1}{2}\Big]\times\Big(
  \int\Big[H(y_1-g(\beta^\top x(F_t),\theta(F_t)))-\frac{1}{2}\Big]\nonumber\\
  &&\times f(z,y_1)dy_1\Big)d F(z,y)+t\int\int\Big[H(y-g(\beta^\top X(F_t),\theta(F_t)))-\frac{1}{2}\Big]\\
  &&\times\Big(
  \int\Big[H(y_1-g(\beta^\top x(F_t),\theta(F_t)))-\frac{1}{2}\Big]\times u(z,y_1)dy_1\Big)d F(z,y)\\
  &&+t\int\int\Big[H(y-g(\beta^\top x(F_t),\theta(F_t)))-\frac{1}{2}\Big]\times\Big(
  \int\Big[H(y_1-g(\beta^\top x(F_t),\theta(F_t)))-\frac{1}{2}\Big]\\
  &&\times f(z,y_1)dy_1\Big)d U(z,y)+t^2\int\int\Big[H(y-g(\beta^\top x(F_t),\theta(F_t)))-\frac{1}{2}\Big]\\
  &&\times\Big(
  \int\Big[H(y_1-g(\beta^\top x(F_t),\theta(F_t)))-\frac{1}{2}\Big]\times u(z,y_1)dy_1\Big)d U(z,y)\\
  &=:&L_1(t)+L_2(t)+L_3(t)+L_4(t).
\end{eqnarray*}
For the first term $L_1(t)$, we have
\begin{eqnarray*}
  \frac{d L_1(t)}{dt}&=&\int\int\frac{d}{dt}H(y-g(\beta^\top x(F_t),\theta(F_t)))\times\Big(
  \int\Big[H(y_1-g(\beta^\top x(F_t),\theta(F_t)))-\frac{1}{2}\Big]\nonumber\\
  &&\times f(z,y_1)dy_1\Big)d F(z,y)+\int\int\Big[H(y-g(\beta^\top x(F_t),\theta(F_t)))-\frac{1}{2}\Big]\\
  &&\times\Big(
  \int\Big(\frac{d}{dt}H(y_1-g(\beta^\top x(F_t),\theta(F_t)))
  \times f(z,y_1)dy_1\Big)d F(z,y).
\end{eqnarray*}
Since $\int[H(y-g(\beta^\top x(F_t),\theta(F_t)))-1/2]d F(y|z)=0$, we have that
$$\frac{d L_1(t)}{dt}|_{t=0}=0.$$
Further,
\begin{eqnarray*}
  \frac{d^2 L_1(t)}{dt^2}|_{t=0}&=&2\int\int\dot{H}(y-g(\beta^\top x(F),\theta(F)))\times\Big(
  \int\Big(\dot{H}(y_1-g(\beta^\top x(F),\theta(F)))\\
  &&\times f(z,y_1)dy_1\Big)d F(z,y).
\end{eqnarray*}
Similarly, it is not difficult to obtain that $\frac{d L_i(t)}{dt}|_{t=0}=0,~i=2,3,4$ and $
\frac{1}{2}\frac{d^2L_2(t)}{dt^2}|_{t=0}~,i=2,3,4$ are equal to other three terms in the formula (\ref{sec_deri}), which completes the proof. $\Box$

\newpage
\begin{table}[h!]
\caption{\linespread{1.15}\small Empirical sizes and powers of $\tilde{S}_n^{OPG}$ and $\tilde{S}_n^{DEE}$ for $H_{10}$ v.s. $H_{11}$ and $H_{12}$ at the significance level $\alpha=0.05$ with $p=8$.
}
\footnotesize
\begin{center}
\begin{tabular}{ccccccccccccccccccccccccc}
\hline
\multicolumn{1}{c}{\multirow{2}{*}{}}&
\multicolumn{1}{c}{\multirow{2}{*}{$\varepsilon$}}&
\multicolumn{1}{c}{\multirow{2}{*}{$a$}}&
\multicolumn{2}{c}{$n=60$}& &\multicolumn{2}{c}{$n=100$}&&
\multicolumn{2}{c}{$n=200$}&
 \\
\cline{4-5} \cline{7-8}\cline{10-11}
\multicolumn{1}{c}{}&\multicolumn{1}{c}{}&\multicolumn{1}{c}{}&
\multicolumn{1}{c}{$\tilde{S}_n^{OPG}$}
&\multicolumn{1}{c}{$\tilde{S}_n^{DEE}$}&&
\multicolumn{1}{c}{$\tilde{S}_n^{OPG}$}
&\multicolumn{1}{c}{$\tilde{S}_n^{DEE}$}&&
\multicolumn{1}{c}{$\tilde{S}_n^{OPG}$}
&\multicolumn{1}{c}{$\tilde{S}_n^{DEE}$}&
\\
\hline
$H_{11}$&$\varepsilon\sim \ln N(0,0.25^2)$
&0&0.048&0.058&&0.049&0.044&&0.050&0.046&\\
&&0.2&0.095&0.127&&0.129&0.133&&0.208&0.163&\\
&&0.4&0.161&0.169&&0.284&0.177&&0.629&0.417&\\
&&0.6&0.306&0.170&&0.562&0.342&&0.925&0.856&\\
&&0.8&0.379&0.255&&0.725&0.553&&0.976&0.972&\\
&&1.0&0.508&0.317&&0.856&0.717&&0.989&0.998&\\
\hline
&$\varepsilon\sim N(0,1)$
&0&0.048&0.057&&0.051&0.049&&0.052&0.047&\\
&&0.2&0.081&0.107&&0.118&0.144&&0.196&0.149&\\
&&0.4&0.164&0.138&&0.267&0.197&&0.614&0.428&\\
&&0.6&0.250&0.174&&0.537&0.347&&0.894&0.837&\\
&&0.8&0.420&0.244&&0.725&0.550&&0.983&0.960&\\
&&1.0&0.479&0.340&&0.850&0.702&&0.991&0.994&\\
\hline
$H_{12}$&$\varepsilon\sim \ln N(0,0.25^2)$
&0&0.045&0.061&&0.045&0.052&&0.052&0.046&\\
&&0.2&0.064&0.103&&0.092&0.142&&0.318&0.372&\\
&&0.4&0.156&0.171&&0.351&0.407&&0.835&0.852&\\
&&0.6&0.296&0.297&&0.606&0.624&&0.976&0.979&\\
&&0.8&0.378&0.406&&0.764&0.753&&0.993&0.995&\\
&&1.0&0.498&0.498&&0.860&0.868&&0.998&0.999&\\
\hline
&$\varepsilon\sim N(0,1)$
&0&0.051&0.059&&0.056&0.055&&0.050&0.046&\\
&&0.2&0.068&0.076&&0.104&0.142&&0.277&0.340&\\
&&0.4&0.136&0.169&&0.350&0351&&0.797&0.829&\\
&&0.6&0.273&0.251&&0.586&0.583&&0.978&0.945&\\
&&0.8&0.377&0.382&&0.759&0.739&&0.997&0.994&\\
&&1.0&0.482&0.473&&0.870&0.859&&0.999&1.000&\\
\hline
\end{tabular}\label{tab_study1}
\end{center}
\end{table}

\newpage

\begin{table}[h!]
\caption{\linespread{1.15}\small Empirical sizes and powers of $\tilde{S}_n^{OPG}$, $\tilde{S}_n^{DEE}$ and $T_n^{WQ}$ for $H_{20}$ v.s. $H_{21}$ and $H_{22}$ at the significance level $\alpha=0.05$.
}
\footnotesize
\begin{center}
\begin{tabular}{ccccccccccccccccccccccccc}
\hline
\multicolumn{1}{c}{\multirow{2}{*}{}}&
\multicolumn{1}{c}{\multirow{2}{*}{$p$}}&
\multicolumn{1}{c}{\multirow{2}{*}{$a$}}&
\multicolumn{3}{c}{$n=100$}& &\multicolumn{3}{c}{$n=200$}&
 \\
\cline{4-6} \cline{8-10}
\multicolumn{1}{c}{}&\multicolumn{1}{c}{}&\multicolumn{1}{c}{}&
\multicolumn{1}{c}{$\tilde{S}_n^{OPG}$}
&\multicolumn{1}{c}{$\tilde{S}_n^{DEE}$}
&\multicolumn{1}{c}{$T_n^{WQ}$}&&
\multicolumn{1}{c}{$\tilde{S}_n^{OPG}$}
&\multicolumn{1}{c}{$\tilde{S}_n^{DEE}$}
&\multicolumn{1}{c}{$T_n^{WQ}$}&
\\
\hline
$H_{21}$&$p=4$
&0&0.045&0.055&0.048&&0.051&0.057&0.045&\\
&&0.2&0.145&0.112&0.055&&0.389&0.164&0.056&\\
&&0.4&0.479&0.273&0.058&&0.871&0.648&0.061&\\
&&0.6&0.708&0.501&0.062&&0.976&0.874&0.067&\\
&&0.8&0.833&0.649&0.067&&0.990&0.956&0.071&\\
&&1.0&0.893&0.733&0.070&&0.997&0.982&0.075&\\
\hline
&$p=2$
&0&0.047&0.057&0.046&&0.053&0.056&0.051&\\
&&0.2&0.165&0.115&0.071&&0.405&0.226&0.129&\\
&&0.4&0.536&0.352&0.121&&0.925&0.681&0.345&\\
&&0.6&0.801&0.527&0.192&&0.986&0.909&0.559&\\
&&0.8&0.882&0.706&0.300&&0.995&0.967&0.698&\\
&&1.0&0.940&0.802&0.344&&0.992&0.989&0.798&\\
\hline
$H_{22}$&$p=4$
&0&0.045&0.049&0.053&&0.047&0.053&0.052&\\
&&0.2&0.105&0.073&0.055&&0.243&0.144&0.056&\\
&&0.4&0.404&0.204&0.058&&0.746&0.496&0.060&\\
&&0.6&0.592&0.379&0.060&&0.932&0.759&0.065&\\
&&0.8&0.771&0.554&0.063&&0.981&0.912&0.070&\\
&&1.0&0.842&0.636&0.067&&0.996&0.956&0.073&\\
\hline
&$p=2$
&0&0.051&0.052&0.059&&0.041&0.055&0.049&\\
&&0.2&0.123&0.097&0.064&&0.275&0.160&0.101&\\
&&0.4&0.429&0.230&0.108&&0.801&0.531&0.252&\\
&&0.6&0.684&0.405&0.170&&0.970&0.800&0.440&\\
&&0.8&0.819&0.597&0.214&&0.994&0.927&0.605&\\
&&1.0&0.897&0.706&0.287&&0.996&0.964&0.717&\\
\hline
\end{tabular}\label{tab_study2}
\end{center}
\end{table}

\clearpage

\begin{table}[h!]
\caption{\linespread{1.15}\small $p$-values of $\tilde{S}_n^{OPG}$ and $\tilde{S}_n^{DEE}$ for the real data analysis.}
\footnotesize
\begin{center}
\begin{tabular}{ccccccccccccccccccccccccc}
\hline
\multicolumn{1}{c}{\multirow{2}{*}{$c_1$}}&
\multicolumn{2}{c}{The first subset}&&\multicolumn{1}{c}{\multirow{2}{*}{$c_2$}}&\multicolumn{2}{c}{The second subset}&
\\
\cline{2-3} \cline{6-7}
\multicolumn{1}{c}{}&
\multicolumn{1}{c}{$\tilde{S}_n^{OPG}$}&$\tilde{S}_n^{DEE}$&
\multicolumn{1}{c}{}& &\multicolumn{1}{c}{$\tilde{S}_n^{OPG}$}&$\tilde{S}_n^{DEE}$&
\\
\hline
3&0.273&0.934&&12&0.802&0.528&\\
4&0.400&0.527&&15&0.587&0.755&\\
5&0.686&0.235&&18&0.374&0.909&\\
6&0.919&0.119&&21&0.207&0.532&\\
\hline
\end{tabular}\label{tab_real}
\end{center}
\end{table}

\end{document}